\documentclass{sigchi}

 \toappear{
  }

\usepackage{balance}       
\usepackage{graphics}      
\usepackage[T1]{fontenc}   
\usepackage{txfonts}
\usepackage{mathptmx}
\usepackage[pdflang={en-US},pdftex]{hyperref}
\usepackage{color}
\usepackage{booktabs}
\usepackage{textcomp}
\usepackage{multicol,graphicx,amssymb,mathrsfs,amsmath, pifont,amscd,latexsym, color,fancyhdr,CJK,url,hyperref,multirow, subfigure, colortbl,tabularx,threeparttable, booktabs, verbatim, bbm,balance, flushend}
\definecolor{mygray}{gray}{.7}
\usepackage{microtype}        
\usepackage{ccicons}          

\usepackage{todonotes}

\usepackage{url}
\usepackage{graphicx, color}
\usepackage{subfigure,caption, comment}
\usepackage[ruled]{algorithm2e}
\usepackage{subfigure,caption}

\usepackage{ragged2e}
\justifying

\newcommand{\tabincell}[2]{\begin{tabular}{@{}#1@{}}#2\end{tabular}}
\def\plaintitle{SIGCHI Conference Proceedings Format}

\def\emptyauthor{}
\def\plainkeywords{Authors' choice; of terms; separated; by
  semicolons; include commas, within terms only; required.}

\makeatletter
\def\url@leostyle{%
  \@ifundefined{selectfont}{
    \def\UrlFont{\sf}
  }{
    \def\UrlFont{\small\bf\ttfamily}
  }}
\makeatother
\def\pprw{8.5in}
\def\pprh{11in}

\definecolor{linkColor}{RGB}{6,125,233}
\hypersetup{%
  pdftitle={\plaintitle},
  pdfauthor={\emptyauthor},
  pdfkeywords={\plainkeywords},
  pdfdisplaydoctitle=true, 
  bookmarksnumbered,
  pdfstartview={FitH},
  colorlinks,
  citecolor=black,
  filecolor=black,
  linkcolor=black,
  urlcolor=linkColor,
  breaklinks=true,
  hypertexnames=false
}

\begin{document}

\title{Mining Device-Specific Apps Usage Patterns from Large-Scale Android Users}


\author{
\alignauthor{Huoran~Li, Xuan~Lu}\\
\affaddr{Peking University}\\
\email{\{lihr, luxuan\}@pku.edu.cn}
}

\maketitle

\begin{abstract}
When smartphones, applications (a.k.a, apps), and app stores have been widely adopted by the billions, an interesting debate emerges: whether and to what extent do device models influence the behaviors of their users? The answer to this question is critical to almost every stakeholder in the smartphone app ecosystem, including app store operators, developers, end-users, and network providers. To approach this question, we collect a longitudinal data set of app usage through a leading Android app store in China, called \textit{Wandoujia}. The data set covers the detailed behavioral profiles of 0.7 million (761,262) unique users who use 500 popular types of Android devices and about 0.2 million (228,144) apps, including their app management activities, daily network access time, and network traffic of apps. We present a comprehensive study on investigating how the choices of device models affect user behaviors such as the adoption of app stores, app selection and abandonment, data plan usage, online time length, the tendency to use paid/free apps, and the preferences to choosing competing apps. Some significant correlations between device models and app usage are derived, leading to important findings on the various user behaviors. For example, users owning different device models have a substantial diversity of selecting competing apps, and users owning lower-end devices spend more money to purchase apps and spend more time under cellular network. 
\end{abstract}


\section{Introduction}

Since Apple announced the iPhones in 2007, smartphones have been playing an indispensable role in people's daily lives. A great variety of applications (a.k.a, apps) such as Web browsers, social network apps, media players, and games make smartphones become the main access channels to Internet-based services rather than communication tools. With the ever-increasing amount of smartphone users and apps, comprehensive and insightful knowledge on what, when, where, and how the apps are used by the users is extremely important~\cite{Alharbi:MobileHCI15}. Many significant research efforts have been made in the past few years on portraying the users and understand their behaviors in term of apps~\cite{TOIS17Liu, TSE17Liu, WWW16Li}.

Like all Internet users, smartphone users can be classified based on various facts, including demographics such as location, gender and age, and behaviors such as preferences to apps~\cite{Chen:WSDM15}, content consumed within apps~\cite{Song:WWW2013}, and so on. Actually, in the current app ecosystem, a user is naturally identified by his/her device~\cite{Xu:IMC11, ICSE16Lu}. Many such classifications of smartphone users boil down to classifications of devices. In other words, much variance of user behaviors may be explained by the devices they use. Indeed, when users interact with their smartphones, download apps from online app stores, and use the apps for different purposes, their experiences are usually affected by various parameters of the device models they use, such as brands, hardware specifications, etc. Understanding how device models affect user behaviors can help app store operators know their users better and improve their recommender systems by considering device models. For instance, one may recommend apps with fancy graphical effects to devices that are equipped with a powerful GPU and high-resolution screen~\cite{khalid2014prioritizing}. Device-specific ads is another big opportunity. For example, Facebook customizes mobile ads according to device model types since 2014~\cite{Web:facebookads}. App developers are thrilled to know through which kind of device models they can gain more users and more clicks, so that they can invest their effort in customizing the ads for those models, e.g., by designing banners of proper sizes or placing videos at proper positions on the screen. Furthermore, device models may be more informative in the behavioral analyses of Android users, due to the large diversity and heavy fragmentation of Android devices~\cite{khalid2014prioritizing}. 

Some existing efforts have been made to investigate how apps usage is affected by device models\footnote{In this paper, the term ``device model" refers to the device with specific product type with hardware specifications, e.g., \texttt{Samsung N7100, N9100}, \texttt{Xiaomi 3s}, and so on}~\cite{Matthias:CHI13, Raptis:MobileHCI13, Zhong:CHI2012, Xu:IMC11}. Unfortunately, due to the lack of sufficient user behavioral data at scale, most existing studies suffer from serious selection bias, including specific user groups (e.g., in-school students)~\cite{Zhong:CHI2012}, fixed device models or apps~\cite{Patro:CoNEXT13}, and limited metrics (e.g., screen size)~\cite{Raptis:MobileHCI13}. In this paper, we present a comprehensive user study exploring whether, how, and how much the device models can really influence the user behaviors on using smartphones and apps. We collect the behavioral profiles of about 0.7 million anonymized Android users\footnote{Our study has been approved by the research ethnics board of the Institute of Software, Peking University. The data is legally used without leaking any sensitive information. The details of user privacy protection are presented later in the data set description. We plan to open the data set when the manuscript is published.} by a leading app store operator in China, called \textit{Wandoujia}\footnote{http://www.wandoujia.com}. Besides the largest data set to date, our study differs from existing efforts further in two aspects. 

\noindent $\bullet$ 
We conduct the study from a new perspective, i.e., \textit{the sensitivity of device's \textbf{price} against the app usage}. In our opinion, the price of a device model can generally reflect the level of hardware specifications when the device is released. Additionally, such a metric can imply the users' economic background, which influences user behavior at demographics level~\cite{Kelly:THMS13}. In this way, we try to categorize the users into different economic groups and explore the sensitivity of the device against the user behavior. 

\noindent $\bullet$ Second, we explore comprehensive behavioral profiles that contain various useful information, including the apps selection, apps management activities (e.g., download, update, and uninstallation), data plan usage per app, and online time length per app, etc. In addition, we focus on only the behavioral profiles from long-term users who steadily contribute to our study. This provides solid ground for the findings from our study. 

The major contributions made by this paper can be summarized as follows.

\begin{itemize}
\item We collect app usage from over 0.7 million users in a period of five months (May 1, 2014 to September 30, 2014). Our data set covers 500 popular Android devices and over 0.2 million Android apps. The detailed user behavioral profiles include the activity log of downloading, updating, and uninstalling apps, the daily traffic and access time of every app through both Wi-Fi and cellular. Based on such a large-scale data set, we explore a comprehensive study on how the device models can impact the app usage.
\item We find significant correlations between the choice of device models and app usage spanning the adoption of app stores, the selection and abandonment of apps, the online access time and data traffic of apps, the revenue of apps, and the preferences against competing apps. Some findings can be quite interesting, e.g., the users holding lower-end devices are likely to spend more money on purchasing apps and spend more time under cellular network, the selection of the apps with similar functionalities presents a substantial diversity among users, etc. The findings can be leveraged to understand the user requirements better, preferences, interests, or even the possible background such as economic or profession.
\item We derive some implications that are directly helpful to several stakeholders in the app-centric ecosystem, e.g., how app store operators can improve their recommendation systems, how the app developers can identify problematic issues and gain more revenues, and how the network service providers can explore more personalized services~\cite{Chinaf14Liu, SOSE13Ma, SOSE13Ma, TSC15Liu, TSC15Huang}. 
\end{itemize}

The remainder of this paper is organized as follows. We first present the related work in the area of user behavior analysis of smartphone users. Next we describe the Wandoujia and the five-month data set, and present our measurement approach alongside the research questions and hypotheses. Then we conduct the correlation analysis on how the choice of device models affects the user behaviors on app usage. In addition, we propose the underlying reasons leading to such significant correlations. We also discuss about our implications for relevant stakeholders, and describe the limitation of our study and threats to the generalization of our results. Finally, we conclude the study and some outlooks to future work.  

\section{Related Work}
Precise classification of users and understanding their behaviors of using apps are significant to every stakeholder in the app ecosystem, including app store operators, content providers, developers, advertisers, network providers, etc. Several efforts have been made in the fields.

One straightforward way to understand the user behavior is conducting field study. Usually, the field studies are conducted over some specific user groups. Rahmati \textit{et al.}~\cite{Rahmati:TMC13, Rahmati:MobileHCI12} made a four-month field study of the usage of smartphone apps of 14 users, and summarized the influences of long-term study and short-term study. Lim \textit{et al.}~\cite{Lim:TSE15} made a questionnaire-based study to discover the diverse usages from about 4,800 users across 15 top GDP countries. The results show that the country differences can make significant impacts on the app store adoption, app selection and abandonment, app review, and so on. Falaki \textit{et al.}~\cite{Falaki:MobiSys10} performed a study of smartphone usage based on detailed traces from 255 volunteers, and found the diversity of users by characterizing user activities. A number of other studies have been made in similar ways~\cite{Matthias:MobileHCI11, Matthias:CHI13, Church:MobileHCI15}. To have more comprehensive behavioral data, the monitoring tools/apps are more appreciated other than questionaire. Yan \textit{et al.}~\cite{Yan:Mobisys11} developed an app, called \textit{AppJoy}, and deployed such an app to collect the usage logs from over 4,000 users and find the possible patterns in selecting and using apps. 

Besides the general analysis, some studies aim to investigate the diversity of user behaviors from specific perspectives. Raptis \textit{et al.}~\cite{Raptis:MobileHCI13} performed a study of how the screen size of smartphones can affect the users' perceived usability,
effectiveness, and efficiency. Rahmati \textit{et al.}~\cite{Rahmati:MobileHCI12} explored how users in different socio-economic status groups adopted new smartphone technologies along with how apps are installed and used. They found that users with relatively low socio-economic background are more likely to buy paid apps. Ma \textit{et al.}~\cite{Ma:WWW2012} proposed an approach for conquering the sparseness of behavior pattern space and thus made it possible to discover similar mobile users with respect to their habits by leveraging behavior pattern mining. Song \textit{et al.}~\cite{Song:WWW2013} presented a log-based study on about 1 million users' search behavior from three different platforms: desktop, mobile, and tablet, and attempted to understand how and to what extent mobile and tablet searchers behave differently compared with desktop users.

Some field studies were made towards specific apps. B{\"{o}}hmer \textit{et al.}~\cite{Matthias:MobileHCI11, Matthias:CHI13} made a field study over three popular apps such as Angry Bird, Facebook, and Kindle. Patro \textit{et al.}~\cite{Patro:CoNEXT13} deployed a multiplayer RPG app game and an education app, respectively, and collected diverse information to understand various factors affecting the app revenues. 

Due to the difficulty of involving a large volume of users, the preceding field studies may suffer from relatively limited users and apps. At large-scale, Xu \textit{et al.}~\cite{Xu:IMC11} presented usage patterns by analyzing IP-level traces of thousands of users from a tier-1 cellular carrier in U.S. They identified traffic from apps based on HTTP signatures and present aggregate results on their spatial and temporal prevalence, locality, and correlation. Our previous work~\cite{Li:IMC15} was conducted over a one-month data collected by Wandoujia, and evidenced that some app usage patterns in terms of app selection, management, network activity, and so on. To understand the diversity of user behaviors better, the study made in this paper is based on a more comprehensive data set that consists of the 5-month behavioral profiles from various users, and focuses on the impact made by the choice of device models.

\begin{table*}[t]
\scriptsize
	\centering
	\caption{Categorization of Device Models}
	\begin{tabular}{c|c|c|c|c}
		\hline
		\textbf{Group} & \textbf{Price Interval} & \textbf{$\#$ of Devices} & \textbf{$\#$ of Users} & \textbf{Representative Devices} \\
		\hline
		\textbf{High-End} & $\geqslant$ 4,000 RMB (about 600 USD) & 77 & 265,636 & \texttt{Samsung N7100, Samsung S4} \\
		\textbf{Middle-End} & 1,000-4,000 RMB (about 150-600 USD) & 339 & 411,138 & \texttt{XIAOMI 3, Google NEXUS 5} \\
		\textbf{Low-End} & $\leqslant$ 1,000 RMB (about 150 USD) & 84 & 84,488 & \texttt{COOLPAD 7231, LENOVO A278T}  \\
		\hline
	\end{tabular}
	\label{tab:user_division}
\end{table*}

\section{The Data Set}
In this section, we present the data set collected from Wandoujia, by describing the detailed information that shall be used to conduct our empirical study. 

\subsection{About Wandoujia}
Our data is from Wandoujia\footnote{Visit its official site via \url{http://www.wandoujia.com}.}, a free Android app marketplace in China. Wandoujia was founded in 2009 and has grown to be a leading Android app marketplace. Like other marketplaces, third-party app developers can upload their apps to Wandoujia and get them published after authenticated. Compared to other marketplaces such as Google Play, apps on Wandoujia are all free, although some apps can still support ``\textit{in-app purchase}". 

\indent Users have two channels to access the Wandoujia marketplace, either from the Web portal, or from the Wandoujia management app. The Wandoujia management app is a native Android app, by which people can manage their apps, e.g., downloading, searching, updating, and uninstalling apps. The logs of these management activities are all automatically recorded. 

Beyond these basic features, the Wandoujia management app is developed with some advanced but optional features that can monitor and optimize a device. These features include network activity statistics, permission monitoring, content recommendation, etc. All features are developed upon Android system APIs and do not require ``root" privilege. Users can decide whether to enable these features. However, these advanced features are supported only in the Chinese version of Wandoujia management app.

\subsection{Data Collection}

Each user of Wandoujia is actually associated with a unique Android device, and each device is allocated a unique anonymous ID. In this study, we collected usage data ranging from May 1, 2014 to September 30, 2014. The data set consists three categories of information: 

\noindent $\bullet$ \textbf{Device Model and Price}. The Wandoujia management app records the type of each user's device model, e.g., \texttt{Samsung Galaxy Note 2, Samsung S3, Xiaomi Note 2}, etc. In our raw data set, there are more than 10,000 different Android device models. Such a fact implies the severe fragmentation of Android devices. However, from our previous work~\cite{Li:IMC15}, the distribution of users against device models typically complies with the \textit{Pareto Principle}, i.e., quite small set of device models accounts for substantial percentage of users. Hence, we choose the top 500 device models according to their number of users. We then look up \textbf{the first-release price} of these device models from \texttt{Jd}\footnote{\url{http://shouji.jd.com}, Jd is the largest e-commerce for electronic devices in China}. Indeed, the price of a device model always decline over time due to the Moore's Law.  However, such a price can somewhat reflect the hardware specifications and the target users at the time when the device model is released on market. Considering the large volume of users, we choose to rely on this type of price as our classification criteria. 

\noindent $\bullet$\textbf{App Management Activities.} The Wandoujia management app can monitor the users' management activities such as searching, downloading, updating, and uninstalling their apps. Each management action is recorded as an entry in the log file that can be uploaded to the Wandoujia servers. With the management activities, we can know which apps ever appeared on the user's device, and when these apps are downloaded, updated, and installed.    
 
\noindent $\bullet$\textbf{App Network Activities.} Besides the basic management functionalities, the Wandoujia management app provides advanced features to record the daily network activities of an app. The network statistic features are optional, and are enabled when and only when the users explicitly grant the permissions to the Wandoujia management app\footnote{Due to space limit, the details of how Wandoujia management app works can be referred to our previous work~\cite{Li:IMC15}}.  If the app generates network connections either from Wi-Fi or cellular (2G/3G/LTE), the network usage can be monitored and recorded at the TCP-level, including the data traffic and network connection time. Note that the network statistic feature works as a system-wide service, thereby network usage from all apps can be captured, even if the app is not installed via the Wandoujia management app. The Wandoujia management app does not record the details of every interaction session, due to the concerns of system overhead. Instead, the Wandoujia management app aggregates the total network usage of an app. In particular, the data traffic and access time generated from foreground and background are distinguished, so that we can have more fine-grained knowledge of network usage. In other words, we can use eight dimensions of the daily network usage per app, i.e., 2 metrics (access time and traffic) * 2 modes (Wi-Fi and Cellular) * 2 states (foreground and background). 


To conduct a comprehensive and longitudinal measurement, we should process and filter the collected data guided by the following principles. We choose only the users who explicitly granted Wandoujia to collect their usage data from the preceding top 500 device models (denoted as the user set $\mathbb{U}$). For these users, we further take into account those who continuously contribute their daily usage data for five months. Such a step assures that we can have rather complete behavioral data of these users. In this way, we obtain \textbf{761,262} users (denoted as the subset $\mathbb{U'}$) and their usage data of \textbf{228,144} apps.


\subsection{User Privacy}
Certainly, the user privacy is a key issue to conduct such a measurement study based on large-scale user behavior data. Besides collecting the network activity data from only the users who explicitly have granted the permission, we take a series of steps to preserve the privacy of involved users in our data set. First, all raw data collected for this study was kept within the Wandoujia data warehouse servers (which live behind a company firewall). Second, our data collection logic and analysis pipelines were completely governed by three Wandoujia employees\footnote{One co-author is the head of Wandoujia product. He supervised the process of data collection and de-identification.} to ensure compliance with the commitments of Wandoujia privacy stated in the \textit{Term-of-Use} statements. Finally, the Wandoujia employees anonymized the user identifiers. The data set includes only the aggregated statistics for the users covered by our study period.

\subsection{Limitation of the Data Set}
The preceding data set may have some limitations. First, the management activities come from only the apps that are operated by the Wandoujia management app. The management activities of apps that are downloaded and updated in other channels such as directly from the app's websites or uninstalled via the default uninstaller of Android system, cannot be captured by our data set. Second, we can take into account only the network usage as an indicator that an app is exactly used. Some apps such as calculator and book readers are often used offline, so we cannot know whether these apps have been launched and how they are used. However, our data set is the largest to date and comprehensive enough to conduct a longitudinal user study.

\begin{table}[t]
\centering
\scriptsize
\caption{Results of U-Test among Groups. The ``H'', ``M'', ``L'' represent High-End, Middle-End, Low-End, respectively.}
\label{tab:t_chi_test}
\begin{tabular}{|c|c|ccc|}
\hline
\multirow{2}{*}{\textbf{RQ}} & \multirow{2}{*}{\textbf{Value}} & \multicolumn{3}{c|}{\textbf{U-Test \textit{p-value}}} \\ \cline{3-5}
                             &                                 & H-M    & H-L    & M-L      \\ \hline
\multirow{2}{*}{RQ1}         & Download \& Update              & 0.000 & 0.000 & 0.000    \\ \cline{2-5}
                             & Uninstallation                  & 0.000 & 0.000 & 0.000    \\ \hline
\multirow{2}{*}{RQ2}         & Cellular                        & 0.000 & 0.000 & 0.000    \\ \cline{2-5}
                             & Wi-Fi                           & 0.000 & 0.000 & 0.000    \\ \hline
\multirow{2}{*}{RQ3}         & Cellular                        & 0.000 & 0.000 & 0.000    \\ \cline{2-5}
                             & Wi-Fi                           & 0.000 & 0.000 & 0.000    \\ \hline
RQ4                          & Expenditure                     & 0.000 & 0.000 & 0.072    \\ \hline
\end{tabular}
\end{table}

\begin{table*}[t]
\centering
\scriptsize
\caption{The Pearson Correlation Coefficient of Each App Category of RQ1-RQ4. The results are presented in form of ``coefficient/p-value''.}
\label{tab:overallcorrelation}
\begin{tabular}{|l|c|c|c|c|c|c|c|}
\hline
\multirow{2}{*}{\textbf{Category}} & \multicolumn{2}{c|}{\textbf{RQ1}} & \multicolumn{2}{c|}{\textbf{RQ2}} & \multicolumn{2}{c|}{\textbf{RQ3}} & \textbf{RQ4} \\ \cline{2-8}
 & \tabincell{c}{\textbf{Download \&} \\\textbf{Update}} & \textbf{Uninstallation}   & \tabincell{c}{\textbf{Cellular} \\\textbf{Time}} & \tabincell{c}{\textbf{Wi-Fi} \\\textbf{Time}} & \tabincell{c}{\textbf{Cellular} \\\textbf{Traffic}} & \tabincell{c}{\textbf{Wi-Fi} \\\textbf{Traffic}} & \textbf{Expenditure} \\ \hline
BEAUTIFY           & -0.252/0.000    & -0.052/0.247 & -0.162/0.000    & -0.148/0.001    & -0.108/0.016       & 0.359/0.000       & -0.086/0.057         \\
COMMUNICATION      & -0.066/0.143    & -0.085/0.059 & 0.205/0.000     & 0.101/0.024     & -0.022/0.624       & 0.302/0.000        & 0.146/0.001          \\
EDUCATION          & -0.150/0.001    & -0.200/0.000 & -0.305/0.000    & -0.079/0.078    & -0.252/0.000       & 0.043/0.337        & 0.077/0.087          \\
FINANCE            & 0.513/0.000     & 0.175/0.000  & 0.127/0.004     & 0.127/0.004     & 0.222/0.000        & 0.129/0.004        & 0.106/0.018          \\
GAME               & -0.567/0.000   & -0.186/0.000 & -0.093/0.038    & -0.024/0.585    & -0.371/0.000      & -0.094/0.036       & -0.221/0.000         \\
IMAGE              & 0.050/0.263     & 0.044/0.323  & 0.093/0.037     & 0.096/0.032     & 0.102/0.023        & 0.239/0.000        & 0.143/0.001          \\
LIFESTYLE          & 0.469/0.000    & 0.139/0.002  & 0.152/0.001     & 0.205/0.000     & 0.322/0.000        & 0.448/0.000        & 0.023/0.609          \\
MOTHER\_AND\_BABY  & 0.080/0.073     & 0.030/0.505  & -0.014/0.757    & -0.002/0.970    & -0.008/0.864       & -0.014/0.764       & -/-                  \\
MUSIC              & -0.407/0.000    & -0.231/0.000 & -0.007/0.885    & -0.041/0.365    & -0.333/0.000       & 0.047/0.290        & -0.019/0.674         \\
NEWS\_AND\_READING & 0.471/0.000     & 0.154/0.001  & 0.230/0.000     & 0.248/0.000     & 0.316/0.000        & 0.394/0.000        & 0.043/0.336          \\
PRODUCTIVITY       & 0.240/0.000     & 0.126/0.005  & 0.298/0.000     & 0.172/0.000     & 0.362/0.000        & 0.312/0.000       & 0.306/0.000         \\
SHOPPING           & 0.488/0.000  & 0.079/0.079  & 0.452/0.000   & 0.402/0.000     &  0.423/0.000      &  0.472/0.000         & 0.016/0.725          \\
SOCIAL             & 0.071/0.115     & 0.143/0.001  & 0.203/0.000     & 0.228/0.000     & 0.188/0.000        & 0.280/0.000        & 0.085/0.057          \\
SPORTS             & 0.094/0.036     & 0.037/0.412  & 0.024/0.587     & 0.024/0.595     & 0.066/0.142        & -0.001/0.991       & -0.131/0.003         \\
SYSTEM\_TOOL       & 0.042/0.352     & -0.037/0.412 & 0.011/0.814     & 0.121/0.007     & 0.072/0.108        &  0.315/0.000      & 0.146/0.001          \\
TOOL               & 0.096/0.031     & 0.055/0.217  & -0.141/0.002    & -0.058/0.198    & -0.234/0.000       & -0.058/0.193       & 0.212/0.000          \\
TRAFFIC            &  0.376/0.000      & 0.049/0.272  & 0.177/0.000     & 0.153/0.001     & 0.238/0.000        & 0.100/0.025        & -/-                  \\
TRAVEL             &  0.560/0.000      &  0.406/0.000   &  0.343/0.000     & 0.200/0.000     & 0.445/0.000      &  0.322/0.000       & 0.049/0.272          \\
VIDEO              & 0.189/0.000     & -0.091/0.042 & 0.066/0.140     & -0.257/0.000    & -0.038/0.403       & -0.369/0.000     & 0.226/0.000          \\
MISCs              & -0.108/0.016    & 0.044/0.328  & -0.007/0.883    & -0.032/0.482    & -0.035/0.433       & 0.216/0.000        & 0.010/0.819          \\
 \hline
\end{tabular}
\end{table*}

\section{Measurement Approach} \label{sec:approach}


In practice, it is a common fact that device models with similar price usually have similar hardware specifications and target user groups. Our measurement study then aims to evaluate whether and to what extent the choice of device models, or more specifically, the price of smartphones, can affect app usage in various metrics. According to the preceding data set of user behavioral profiles, we propose some research questions that are significant to stakeholders.

\begin{itemize}
\item \textbf{RQ1:} Does the choice of device models impact the usage of app stores? If such an impact exists, which users are more likely to adopt the app stores (e.g., downloading new apps and updating existing apps), and what about the various requirements of different users when they use app stores?  
\item \textbf{RQ2:} Does the choice of device models impact the online time spent by users? If such an impact exists, which apps do different users tend to spend their time on?
\item \textbf{RQ3:} Does the choice of device models impact the traffic used by users, especially the data plan of cellular? If such an impact exists, which apps are more likely to consume the cellular data plan of different users?
\item \textbf{RQ4:} Does the choice of device models impact the purchase behavior of users and thus the app revenue? If such an impact exists, which group of users are more likely to pay for apps, and which apps are more likely to be paid for by different users?
\item \textbf{RQ5:} Does the choice of device models impact the choice of apps with similar functionalities or purposes? If such an impact exists, which apps are more likely to be adopted by different users?
\end{itemize}

To answer the above \textbf{RQ}s, our measurement study is conducted from two aspects. 
\subsubsection{User Group Analysis}
As we assume that the price of device models can possibly reflect the economic background of the users, we first study the \textbf{overall} user behaviors by categorizing the users into groups according to the first-release price. In China, the price systems of popular e-commerce web sites such as \texttt{Jd}, \texttt{Amazon}, and \texttt{Taobao}, the price of device models is usually segmented at every 1,000 RMB level, i.e.,  $\leqslant$ 1,000 RMB, 1,000 RMB-2,000 RMB, 2,000 RMB-3,000 RMB, 3,000 RMB-4,000 RMB, and $\geqslant$4,000 RMB. Hence, we roughly categorize the device models into three groups according to their on-sale price information that is published on the \texttt{Jd}, i.e., the \textbf{High-End} ($\geqslant$4,000 RMB, about 600 USD), the \textbf{Middle-End} (1,000 RMB-4,000 RMB, about 150-600 USD), and the \textbf{Low-End} ($\leqslant$ 1,000 RMB, about 150 USD). We list the categorization results in Table~\ref{tab:user_division}. 

We manually check the price history evolution of the top 500 device models on \texttt{Jd} as well as look up some third-party data sources such as \texttt{Dong-Dong}\footnote{\url{https://itunes.apple.com/us/app/dong-dong-gou-wu-zhu-shou/id868597002?mt=8}, is an app for inquiring history price of products on \texttt{Jd}.} and \texttt{Xitie}\footnote{\url{http://www.xitie.com}, is a website for inquiring price history of products on popular e-commerce sites.}. Most of the device models were first released to market after 2012, and can still fall into the above coarse-grained groups as of May 1, 2014 (the start time of our data set). Very few device models cannot meet such criteria, e.g., the first-release price of \texttt{Galaxy S2} was 4,399 RMB, but the price fell down to about 3,200 RMB as of May 1, 2014. For this case, we still categorize the device models by the first-release price. Luckily, such  is rare in our data set. In this way, the dynamics of prices can have the side effects as little as possible. For each \textbf{RQ}, we use the box-plot to report the distribution of user behaviors over the corresponding metrics in each group. To further evaluate the statistical significance among the groups, we employ the Mann-Whitney U test (U-Test for short)~\cite{book:ISMDA} among the three groups. The U-Test is widely adopted in large data set to test whether the two groups have significant difference. In particular, the U-Test can be applied onto unknown distributions, and fit our data set very well. 

\subsubsection{App Category Analysis}
We then study whether the choice of device models can impact the user behavior on specific apps. To this end, we run the \textit{regression analysis} between the price of device model and the corresponding metrics, by organizing the $\mathbb{U'}$ according to the first-release price of each device model. Investing each \textbf{RQ} essentially relates to the user's preferences and requirements of specific apps. Hence, we categorize the apps according to the classification system of Wandoujia, e.g., \textit{NEWS\_AND\_READER}, \textit{GAME}, \textit{VIDEO}, and so on. For each \textbf{RQ}, we summarize the related metrics over each app category for every single user, and thus compute the Pearson correlation coefficient between the metrics and the price of the device model. In this way, we can explore whether the choice of device models can be statistically significant to the app usage.

%

In the next section, we conduct our analysis in the following workflow. We first present the motivations of each research question, respectively. Then we try to examine the impact caused by the choice of device models,by synthesizing the correlation analysis results at the granularity of each device model and the groups. Finally, we summarize the findings and try to explore the underlying reasons leading to the diverse user behaviors on app usage.
\section{Analysis and Results}
In this section, we explore the research questions and validate all the hypotheses, respectively. We show the results of user group U-Test in Table~\ref{tab:t_chi_test} and the results of app-specific regression analysis in Table~\ref{tab:overallcorrelation}. 

Generally, most of the pairwise U-Test results in Table~\ref{tab:t_chi_test} at \textit{p} < 0.0001 level. The only exception is for \textbf{RQ4}, indicating that the difference on expenditure of paid apps is not quite significant between middle-end group and low-end group. Such an observation indicates that the \textbf{price of device model has a statistically significant correlation of the behaviors of users from different groups}. Hence, we can focus more on the distribution variance among groups (in the box-plots) and the results reported in Table~\ref{tab:overallcorrelation}. 


\subsection{Effect on App Management Activities}

\begin{figure}
	\begin{center}
	\subfigure[Download \& Update\label{fig:user_dp}]
	{
		\includegraphics[width=0.22\textwidth]{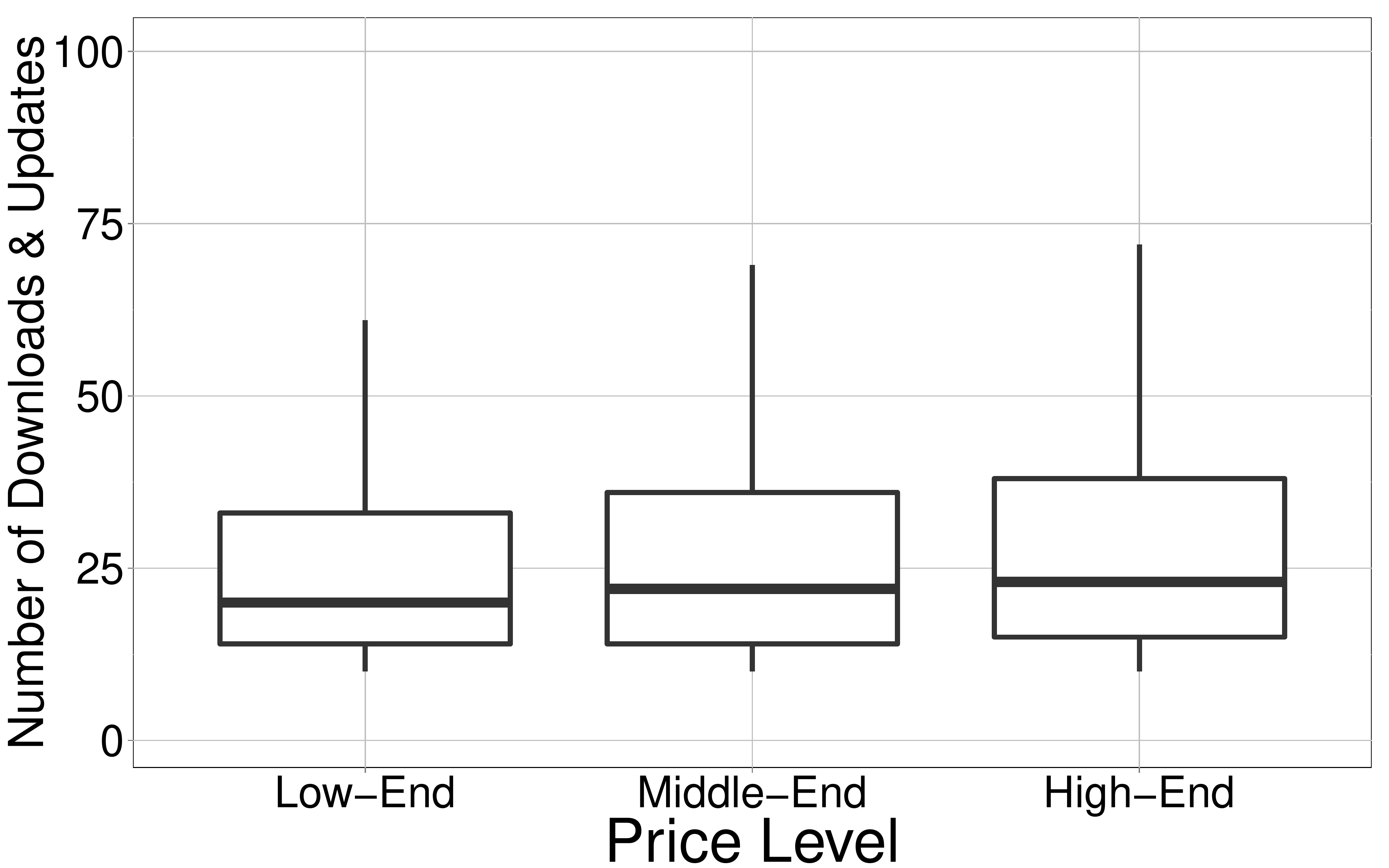}
	}
	\subfigure[Uninstallation\label{fig:user_u}]
	{
		\includegraphics[width=0.22\textwidth]{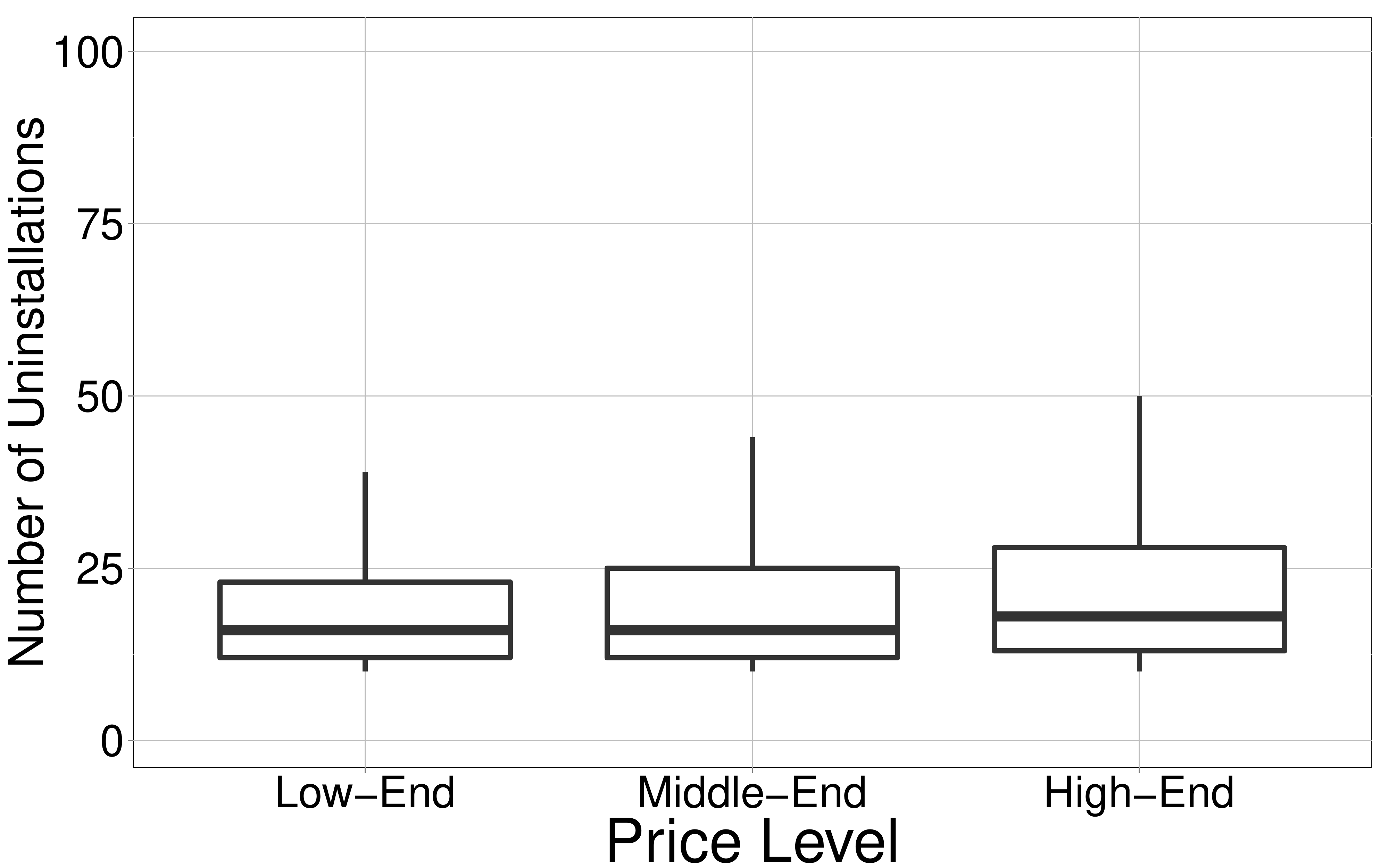}
	}
	\caption[7.5pt]{Distributions of the Number of Downloads, Updates, and Uninstallations among Device Model}\label{fig:group_management}
	\end{center}
\end{figure}

First, we focus on \textbf{RQ1}, i.e., whether users holding different device models perform differently in terms of app downloads, updates, and uninstallations. This research question is motivated by three folds. First, the behaviors of app downloads can indicate the adoption of app stores by users, i.e., whether the users prefer seeking apps from app stores and which users are more active. Second, we can identify the most popular/unpopular apps for a given device model, the app stores operators can accurately recommend the apps. Third, we can explore which apps are more likely to be abandoned by the users holding a specific device model, and such knowledge can help app developers identify the possible problems such as device-specific bugs.

\subsubsection{Adoption of App Stores}
In Figure~\ref{fig:group_management}, we first report the number of 5-month management activities in terms of download and update. On average, we can see that most users do not frequently access the app stores. However, in each group, the standard deviation of management activities is quite significant. Such an observation indicates that users can perform quite variously in terms of management. We can observe that the users holding higher-level devices are more likely to access the app stores.



\subsubsection{App Selection}
We next investigate whether the choice of device models can impact the app selection. From our previous study of the global distribution of apps~\cite{Li:IMC15}, we find that users can have quite high overlap in selecting the popular apps, such as \texttt{WeChat}, \texttt{QQ}, \texttt{Map}. Hence, we first explore the \textbf{similarity} of apps selection. We aggregate the apps that are downloaded and updated by at least 1,00 unique users of each device model, and compute the pairwise cosine similarity between the three groups. The cosine similarity values are 0.81 (high-end and middle-end), 0.86 (high-end and low-end), and 0.81 (middle-end and low-end), respectively. Such an observation evidences our preliminary findings. Then we explore the \textbf{diversity} of app selection. For simplicity, we cluster each app according to its category information provided by Wandoujia, e.g., \textit{Game}, \textit{NEWS\_AND\_READING}, etc. We compute the contributions of downloads and updates from every single device model against a specific app category. For example, if there are 1,000,000 downloads of \textit{GAME} apps and 50,000 of these downloads and updates come from the device model \texttt{Samsung S4}, we assign the contributions made by this device model is 5\%. Then we make the correlation analysis of app selection and the price of device, by means of Pearson correlation co-efficient. We find that as the price of device models increases, the users are more likely to choose apps from the categories of \textit{TRAFFIC} (\textit{r}=0.376, \textit{p} = 0.000) , \textit{LIFESTYLE}  (\textit{r} = 0.469, \textit{p} = 0.000), \textit{NEWS\_AND\_READING}  (\textit{r} = 0.471, \textit{p} = 0.000), \textit{SHOPPING}  (\textit{r} = 0.488, \textit{p} = 0.000), \textit{FINANCE}  (\textit{r} = 0.513, \textit{p} = 0.000), and \textit{TRAVEL}  (\textit{r} = 0.560, \textit{p} = 0.000). In contrast, the correlation analysis show that as the prices of device models increases, the users are less likely to choose the apps from \textit{GAME} (\textit{r} = -0.567, \textit{p} = 0.000) and \textit{MUSIC} (\textit{r} = -0.407, \textit{p} = 0.000).  Such observations imply that the choice of device models can significantly impact the app selections, and infer the characteristics and requirements of the users. For example, users with high-end smartphones are more likely to have higher positions and better economic background, so they care about the apps from \textit{NEWS AND READING}, \textit{FINANCE}, \textit{TRAVEL}, and \textit{SHOPPING}. Users holding low-end device models care more about the entertainment such as \textit{Game} and \textit{Music}.

\begin{figure}
	\begin{center}
	\subfigure[Cellular\label{fig:time_cellular_eachgroup}]
	{
		\includegraphics[width=0.22\textwidth]{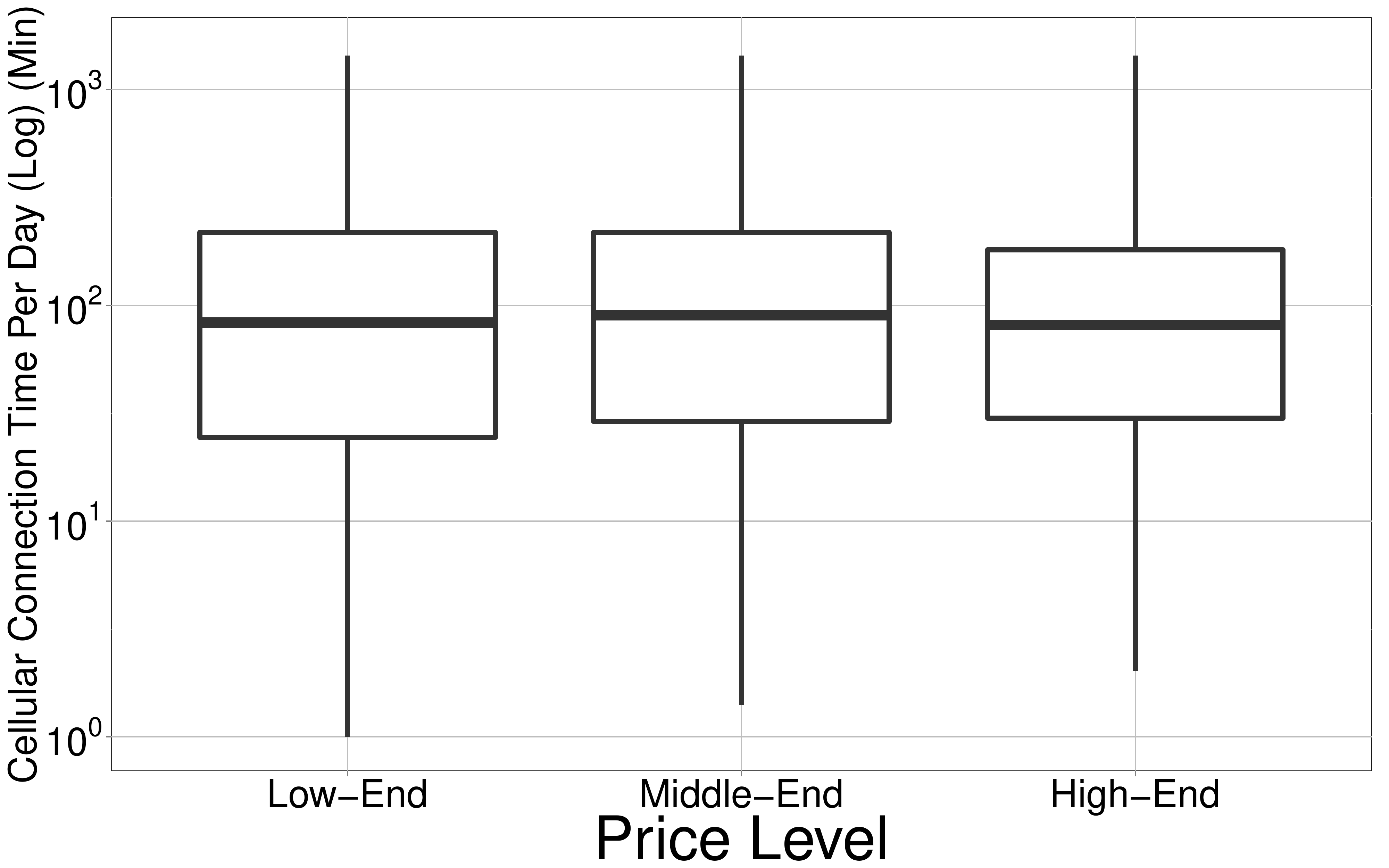}
	}
	\subfigure[Wi-Fi\label{fig:time_wifi_eachgroup}]
	{
		\includegraphics[width=0.22\textwidth]{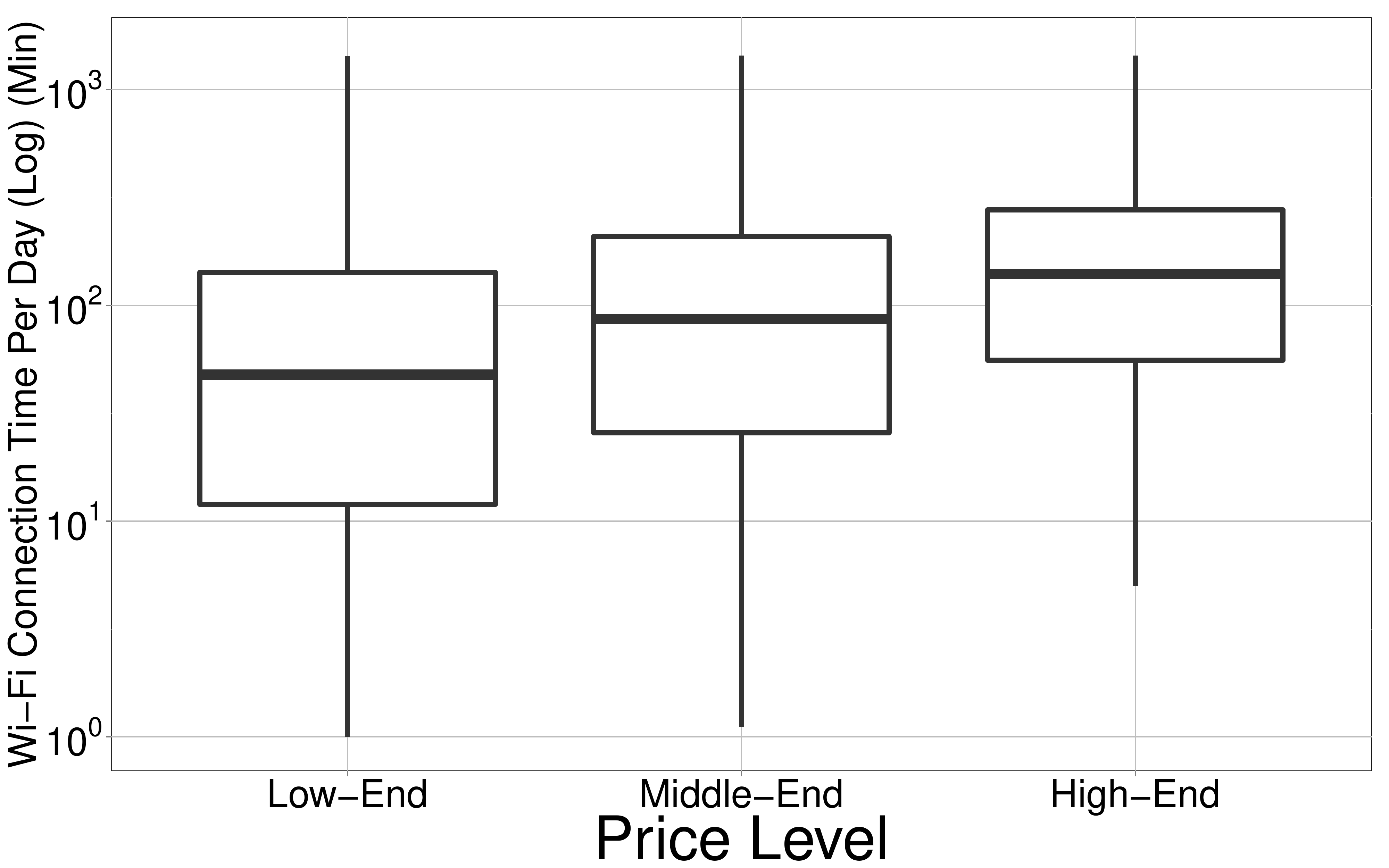}
	}
	\caption[7.5pt]{Distribution of Daily Online Time among Device Models.}\label{fig:time_eachgroup}
	\end{center}
\end{figure}

\subsubsection{App Abandonment} 

The uninstallation can indicate the users' negative attitudes towards an app, i.e., the user does not like or require the app any longer. From Figure~\ref{fig:group_management}, it is observed that the median of uninstallation activities differ quite marginally among the three groups. We then perform the correlation analysis in a similar way of downloads and updates. In most categories, the Pearson co-efficient is not quite significant. However, we find that the correlation in \textit{TRAVEL} (\textit{r} = 0.406, \textit{p} = 0.000) is significant. Although the travel apps are most likely to be downloaded by higher-end users, these apps are also very possible to be uninstalled by these users. 

Although the uninstallation does not take significant correlations to device models at the level of app category, investigating the individual apps that are possibly abandoned by a specific device model is still meaningful. To this end, we explore the apps which have been uninstalled for more than 100 times in our data set, and get 3,123 apps. We then draw the distribution of uninstallations according to the device model. An interesting finding is that, the manufacturer-customized or preloaded apps are more possibly uninstalled on their own lower-end device models. For example, the app \texttt{Huawei Cloud Disk} (the package \texttt{com.huawei.hidisk}) is a preloaded app on almost all device models produced by \texttt{Huawei}. This app has 20,985 uninstallations, while 17,641 uninstallations come from the lower-end devices. The similar findings can be found in other device models produced by \texttt{Samsung, Lenovo}, and \texttt{ZTE}. Such an observation implies that the lower-end users are less likely to adopt these customized or preloaded apps. Besides the preloaded apps, some apps are also more likely to be uninstalled by a specific device model. For example, the \texttt{Samsung Galaxy Note 2} accounts for more than 80\% uninstallations of two camera apps. Such a finding implies that these apps can suffer from device-specific incompatibility or bugs.

\subsection{Effect on Online Time}
We next intend to validate \textbf{RQ2}, i.e., the choice of device models impacts on how long the users spend their time on the Internet. Such a research question is motivated by understanding whether the choice of device models can lead to various preferences toward cellular and Wi-Fi usage. If the app developers can know that some users from a specific device model spend more cellular time rather than Wi-Fi, they can provide customized features to these models. For example, developers can optimize the data plan usage by providing pre-downloading contents when these users are in Wi-Fi network. In addition, if users holding specific device models spend much time on a few categories of apps, the developers, web content providers, and advertisers can leverage such knowledge to customize more accurate advertisements to audience.

The Wandoujia management app can record the daily foreground/background connection time under both cellular and Wi-Fi. Since the foreground time is computed only when the users interact with the app (by checking the stack of active apps in Android system), we exclude the background time in the online time analysis. 


Figure~\ref{fig:time_eachgroup} describes the distribution of online time. \textbf{For online time, we are surprised to find that users rely less on the cellular network as the price of device model increases. } In other words, the higher-end users typically spend less time under cellular network. For the average daily online time, the low-end users ($\leqslant$ 1,000-RMB device models) spend about 10 minutes more than the high-end users ($\geqslant$ 4,000-RMB device models) under cellular, while the high-end users spend 1 hour more than the low-end users under Wi-Fi. Immediately, we can infer that the network conditions vary a lot among different users, i.e., the lower-end users are less likely to stay in the places where Wi-Fi are covered. In contrast, the higher-end users tend to have better Wi-Fi connections. Such a difference can further imply the possible locations where different users may stay, e.g., the high-end users are more likely to stay in offices. 

We then investigate whether the choice of device models affect the usage of ``network-intensive" apps. Similar to the preceding analysis in the management activities, we compare the online time distribution of device models over each app category, under cellular and Wi-Fi, respectively. For cellular, the online time of apps have no significant correlation with the price of device models, except the categories of \textit{TRAVEL} (\textit{r} = 0.3427, \textit{p} = 0.000) \textit{SHOPPING} (\textit{r} = 0.452, \textit{p} = 0.000), and \textit{EDUCATION} (\textit{r} = -0.305, \textit{p} =0.000).

This result can support the findings of \textbf{RQ1}. On the other hand, it is interesting to see that users holding lower-end smartphones are more likely to use \textit{EDUCATION} (\textit{r} = -0.305, \textit{p} = 0.000) apps under the cellular. Such a finding suggests that a considerable proportion of lower-end users may be students. 
The correlation between the choice of device model and online time under Wi-Fi seems not to be quite significant, either. Only in the category of \textit{SHOPPING} (\textit{r} = -0.304, \textit{p} = 0.000), the choice of device models seems to take significant correlation with the price of device. Such an observation is not surprising, as higher-end users are supposed to have better economic background and more likely to spend more on shopping. We can infer that users share quite similar preferences in app usage under Wi-Fi.

%

\subsection{Effect on Traffic Consumption}

\begin{figure}
	\begin{center}
	\subfigure[Cellular\label{fig:traffic_cellular_eachgroup}]
	{
		\includegraphics[width=0.22\textwidth]{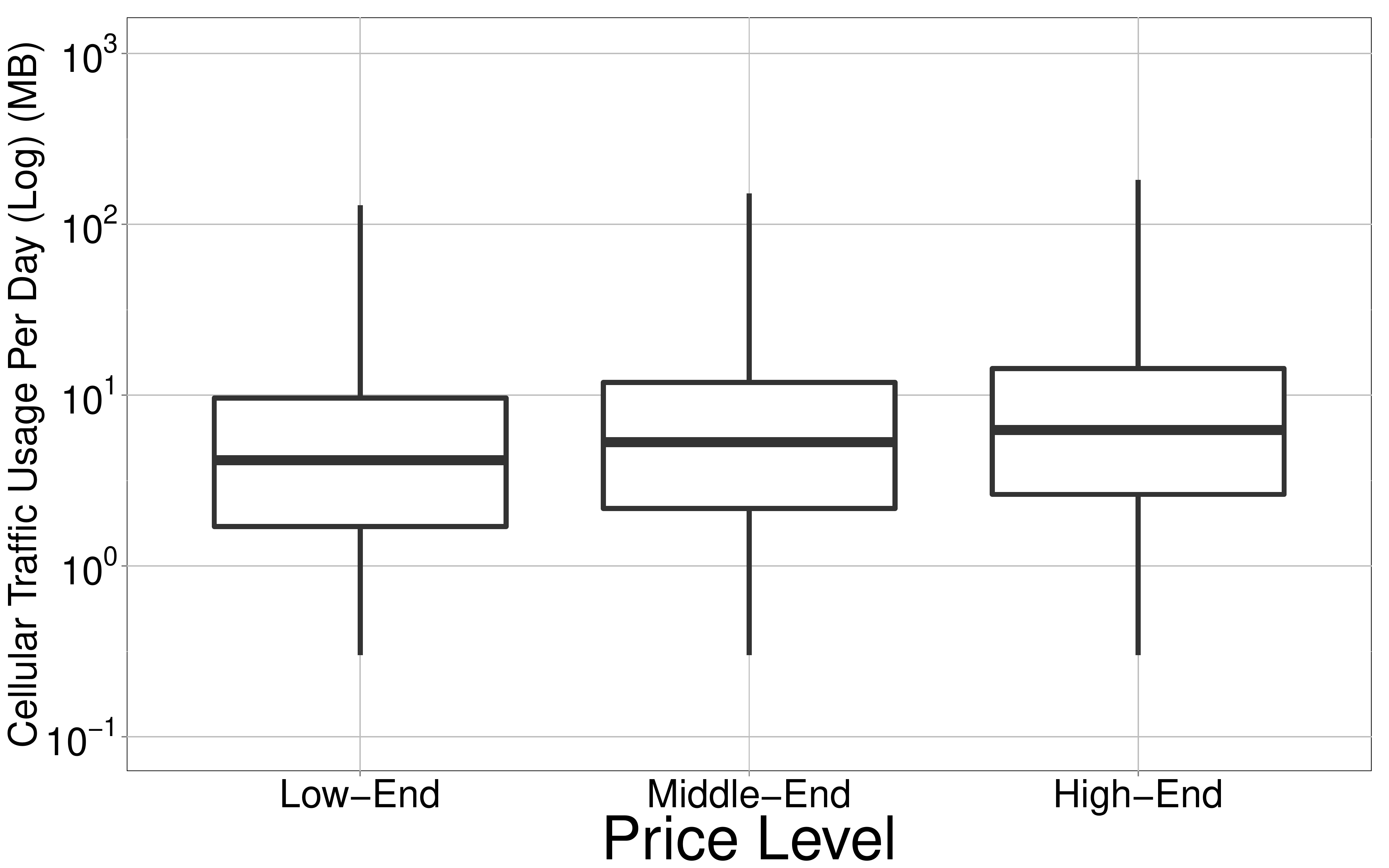}
	}
	\subfigure[Wi-Fi\label{fig:traffic_wifi_eachgroup}]
	{
		\includegraphics[width=0.22\textwidth]{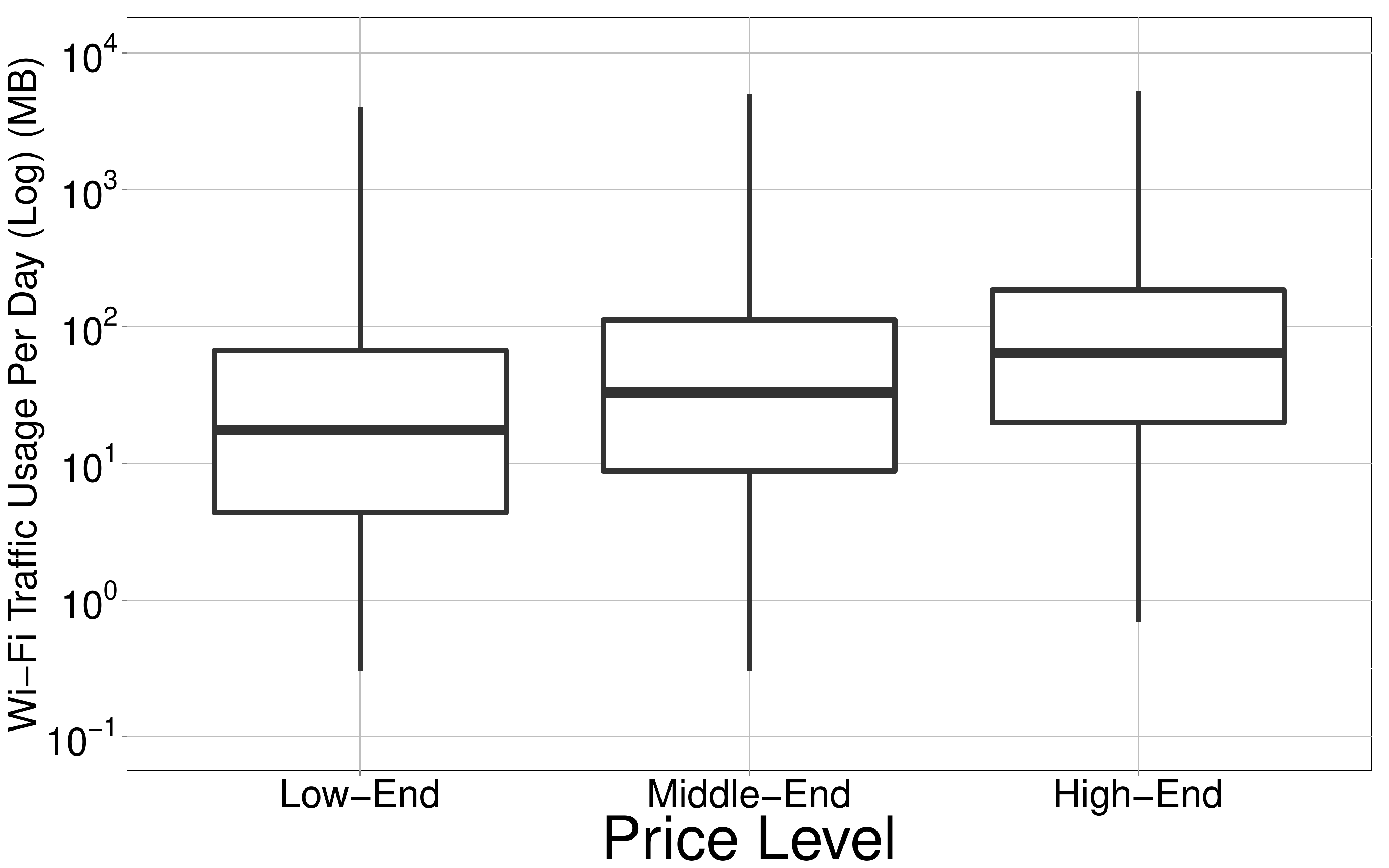}
	}
	\caption[7.5pt]{Distribution of Daily Traffic Consumption among Device Models.}\label{fig:traffic_eachgroup}
	\end{center}
\end{figure}

Another important network-related issue for smartphone users is the traffic. We next focus on \textbf{RQ3}, i.e., whether users holding different device models differ in cellular and Wi-Fi traffic usage. Generally, users do not care about the traffic from Wi-Fi, but do concern the traffic from cellular that they need to pay for, especially for low-end users who usually have relatively limited budgets. However, exploring the traffic generated from Wi-Fi can be meaningful. Intuitively, we can explore which apps are more ``bandwidth-sensitive" on specific device models, so the app developers and network service provider can consider techniques to compensate for bandwidth variability.  

The distribution of traffic consumption among device models is shown in Figure~\ref{fig:traffic_eachgroup}. Interestingly, although the higher-end users are observed to spend the least time under the cellular network in \textbf{RQ2}, they spend the most traffic. In other words, we can infer that the higher-end users are more likely to use those ``traffic-heavy" apps. On average, a high-end user can spend more 100 MB cellular data plan than a low-end user. In China, such a difference of data plan does matter very much. 

Since the gaps between different device models are substantial, identifying which apps consume most traffic on specific device models can be quite meaningful. Similar to preceding analysis, we compute the Pearson correlation coefficients between the choice of device model from every single user and the apps on which the traffic is consumed. The cellular data plan consumed over the apps from \textit{SHOPPING} (\textit{r} = 0.423, \textit{p} = 0.000) and \textit{TRAVEL} (\textit{r} = 0.445, \textit{p} = 0.000) presents a quite significant positive correlation to the price of device models. In contrast, the correlations seem to be significantly negative in \textit{GAME} (\textit{r} = -0.371, \textit{p} = 0.000) and \textit{MUSIC} (\textit{r} = -0.333, \textit{p} = 0.000) apps. In these app categories, users with lower-end smartphones tend to spend more cellular traffic. These findings are consistent with the app download and update preferences in \textbf{RQ1}. 

The traffic generated under Wi-Fi presents significant correlations with the device models in some categories. The lower-end users tend to spend a large number of Wi-Fi traffic on the \textit{VIDEO} apps, In contrast, the higher-end users are more likely to rely on the apps of \textit{COMMUNICATION}, \textit{PRODUCTIVITY}, \textit{SYSTEM\_TOOL}, \textit{TRAVEL}, \textit{BEAUTIFY}, \textit{NEWS\_AND\_READING}, \textit{LIFESTYLE}, and \textit{SHOPPING} under Wi-Fi. Such a difference in the Wi-Fi traffic usage can indicate the requirements and preferences of users holding different device models, and thus implies the possibly different background of the users. 


\subsection{Effect on App Revenue}

The above three research questions have revealed the general correlations between the choice of device models and app usage. To further explore the underlying reasons that can be more relevant to other stakeholders such as app developers, we try to focus on the \textbf{RQ4}, i.e., whether the choice of device models can affect the app revenue. It should be mentioned that all apps directly downloaded from Wandoujia are free, despite the in-app purchase of some apps. However, since the Wandoujia management app act as a system-wide service to monitor all apps that are installed on a device, we can still identify which apps should be paid ones that are downloaded from other channels. To this end, we write a crawler program to retrieve the package names of paid Android apps from other app stores including Google Play, 360safe App Store, and Baidu App Store. Then we extract the logs that are related to these paid apps from our data set. In addition, we record the fee of each paid app. By such a step, we finally obtained 27,375 users that have used paid apps.  
As the price of a device model can serve as an indicator of its owner's economic background, it is not uncommon to suppose users with expensive devices to pay more money in purchasing paid apps. We are interested in three facets, (1) \textbf{\textit{the possibility to buy paid apps}}, i.e., are users with high-end devices more likely to use paid apps? (2) \textbf{\textit{the expenditure of apps}}, i.e., do the high-end users pay more money on buying apps? (3) \textbf{\textit{the diversity of paid apps}}, i.e., do users with different models pay for different purposes? 

First, we investigate the possibility of buying apps. From Table~\ref{tab:npurchase}, we can observe that the higher-end users account for about 47\% of all users who have ever purchased paid apps, and 45\% of purchased records. We further perform a \textit{Chi-square test} to confirm the significant correlations between the proportion of buyers and group ($\chi^2=2875.05$, \textit{p} = 0.000).  

When moving to the expense of apps, we perform the pairwise T-test of average expenditure of users. The T-Test can identify the positive or negative correlation of the tendency of buying apps against the groups. The T-test values are -4.281 (high-end v.s. low-end) and -5.740 (middle-end v.s. low-end), respectively, given the \textit{p} = 0.000. To our surprise, buyers with higher-end device models are likely to spend even less money than lower-end users. Interestingly, \textbf{although the lower-end users account for quite small proportion of all buyers and the purchase orders, they are likely to spend more money on average.} Such a finding indicates that the choice of device models impact the wish to buy apps, which affects the revenue of apps accordingly.

Finally, the choice of device models can impact the preferences of paid apps. The lower-end users are more likely to pay for \textit{GAME} (\textit{r} = -0.221, \textit{p} = 0.000), \textit{BEAUTIFY} (\textit{r} = -0.086, \textit{p} = 0.000 such as themes), and \textit{SPORTS}  (\textit{r} = -0.131, \textit{p} = 0.000). In contrast, the higher-end users tend to buy \textit{PRODUCTIVITY} (\textit{r} = 0.306, \textit{p} = 0.000, such as office suite). Such a difference can reflect the different requirements of apps which the users would like to pay for.


\begin{table}[hbp]
	\centering
	\scriptsize
	\caption{The Number of Paid Apps, Buyers, and Purchase Records}
	\label{tab:npurchase}
	\begin{tabular}{|c|c|c|c|}
	\hline
	\textbf{Device Group} & \tabincell{c}{\textbf{\# of}\\\textbf{Paid apps}} & \tabincell{c}{\textbf{\# of}\\\textbf{Buyers}} & \tabincell{c}{\textbf{\# of}\\\textbf{Purchases}} \\ \hline
	High-End       & 639                           & 13,788                & 14,977                         \\
	Middle-End     & 628                           & 13,700                & 14,591                         \\
	Low-End        & 190                           & 1,590                 & 1,640                          \\ \hline
	\end{tabular}
\end{table}

%

\subsection{Effect on Choices of Competing Apps}

We finally move to the \textbf{RQ5}, i.e., whether the choice of device models can affect selecting the apps of the same or similar functionalities (we name such apps as ``competing apps" in the follows) Such a research question is motivated by the existence of a number of ``competing apps" on the app stores. For example, there are a number of competing browsers such as \texttt{Chrome, FireFox, Opera Mini, Safari}, maps such as \texttt{Google Maps, Baidu Maps, and Yahoo! Maps}, and so on. Although these apps can provide very similar or even the same functionalities, they can perform quite variously such as the data traffic and energy drain, given the same user requests~\cite{Zhong:SIGMobileComm13}. In addition, besides the common functionalities, the competing apps are likely to provide some differentiated features such as adjustable color and light for display. End-users often feel confusing to select the apps that are more adequate to their own preferences and requirements.  

Unlike the correlation analysis of \textbf{RQ1-RQ4}, we do not directly conduct correlation analysis to all app categories. Instead, we choose three typical apps: \textit{News reader}, \textit{Video player}, and \textit{Browser}, as they are observed to be commonly used in daily life. For each app, we select the top-5 apps according to the online time that the users spend on them. The reason why we employ the online time instead of the number of downloads is that the online time can be computed only when the users interact with the app. The selected competing apps are as follows. The \textbf{News} contains \texttt{Phoenix News, Sohu News, Netease News, Today's Top News}, and \texttt{Tencent News}; the \textbf{Video Player} contains \texttt{QVOD, Lenovo Video, Baidu Video, Sohu Video}, and \texttt{iQiyi Video}; the \textbf{Browser} contains \texttt{Chrome, UC Web, Jinshan, Baidu}, and \texttt{360Safe}.

\begin{figure*}
	\begin{center}
	\subfigure[News reader\label{fig:competing_cdf_news}]
	{
		\includegraphics[width=0.28\textwidth]{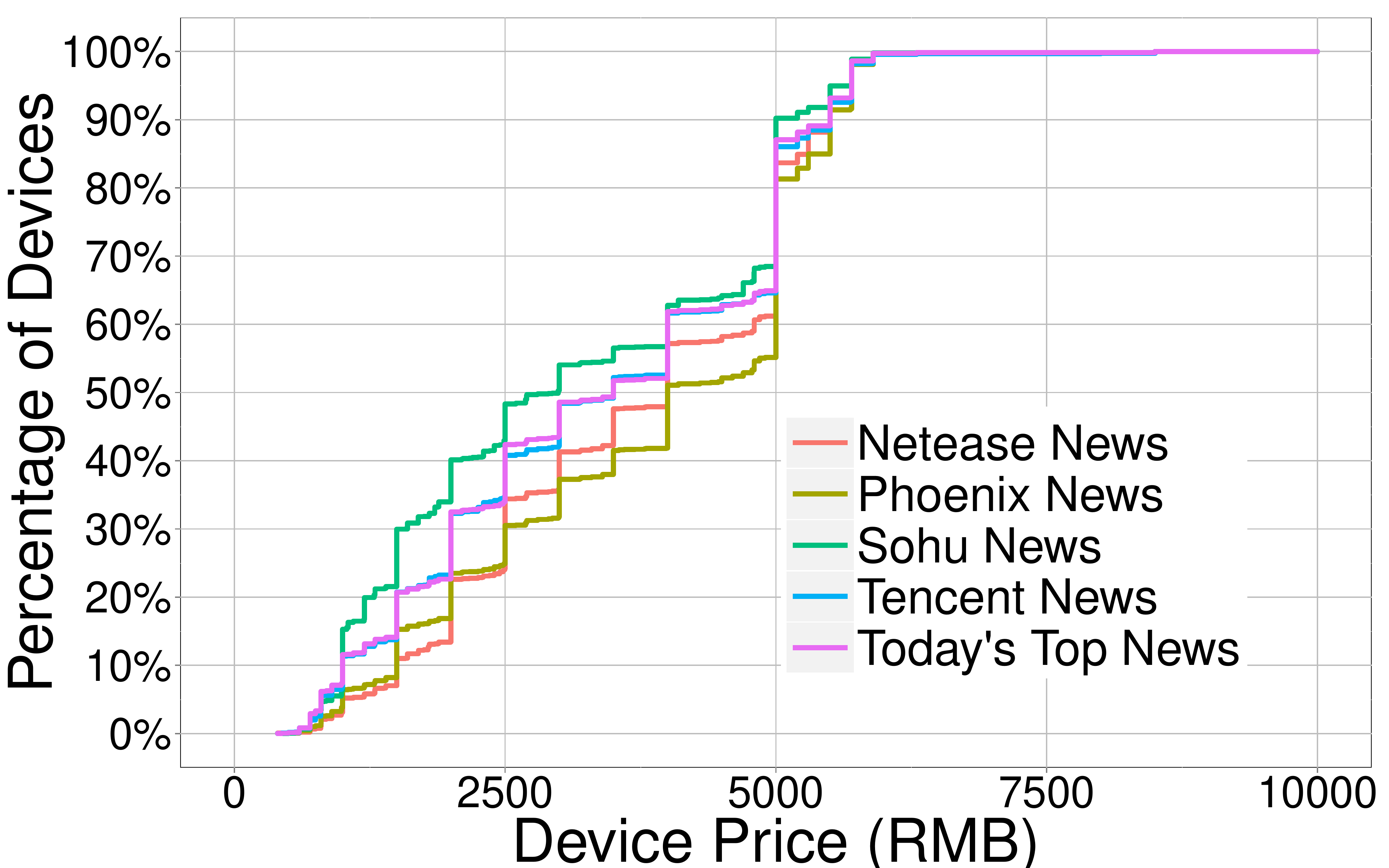}
	}
	\subfigure[Video player\label{fig:competing_cdf_video}]
	{
		\includegraphics[width=0.28\textwidth]{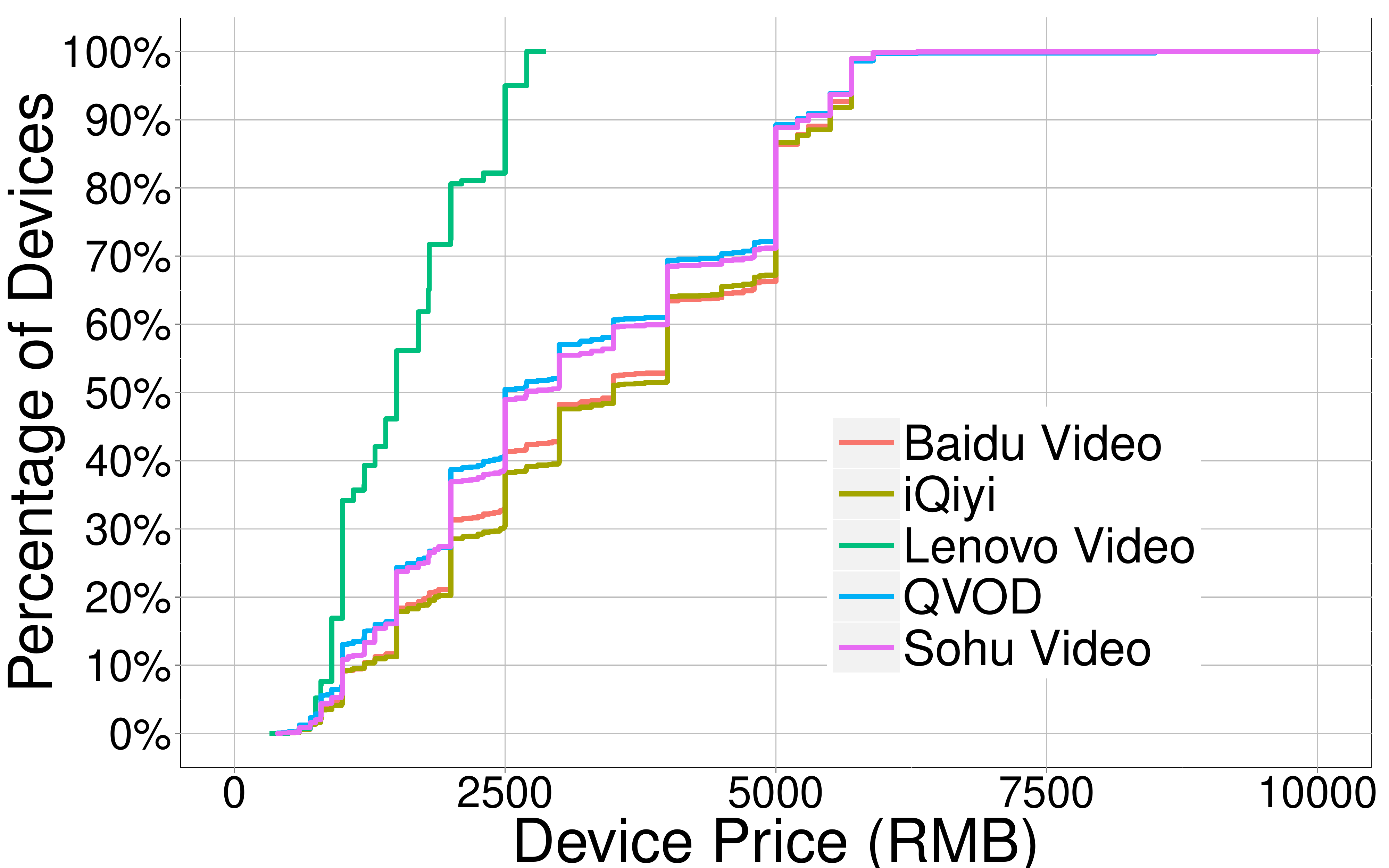}
	}
	\subfigure[Browser\label{fig:competing_cdf_browser}]
	{
		\includegraphics[width=0.28\textwidth]{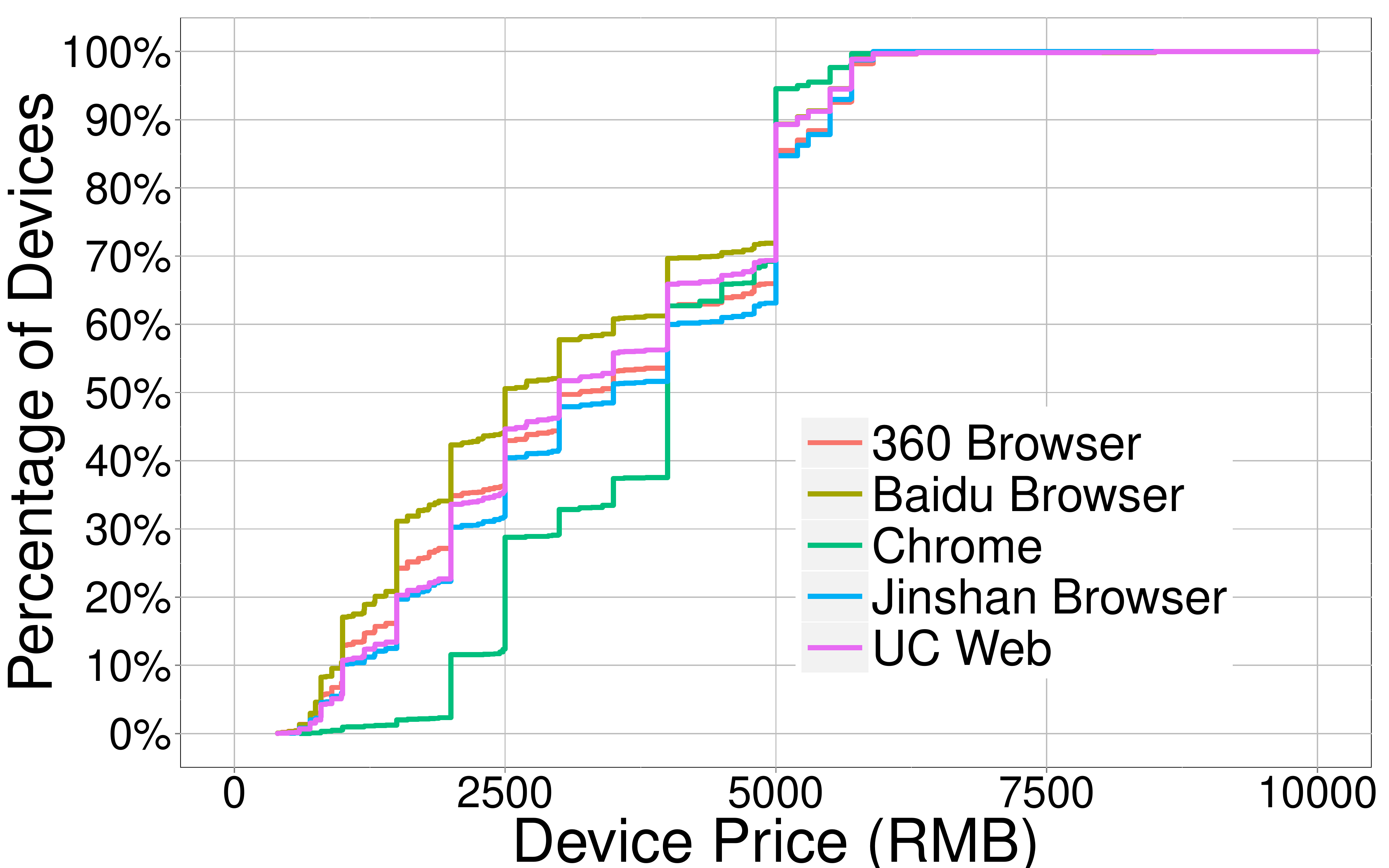}
	}
	\caption[7.5pt]{Distribution of Preferences of Competing Apps}\label{fig:competing_cdf}
	\end{center}
\end{figure*}

First, we want to figure out the distribution of the user preferences against the app according to the device model. We employ the cumulative distribution function (CDF) to demonstrate such distributions, as shown in Figure~\ref{fig:competing_cdf}. For each app, the X-Axis represents the price of device models that are sorted in ascending order, and the Y-Axis refers to the percentage of the app's users holding such a device model. An app tends to have be used by more higher-end users if the curve draws near the bottom. Obviously, we can observe that the choice of device models significantly impact the selection of competing apps. For the news readers, we can see that the \texttt{Phoenix News} and \texttt{Netease News} are more likely to be adopted by higher-end users. In contrast, the \texttt{Sohu News} tend to be more preferred by the lower-end users. 

The difference among device models is even more significant for the \textit{Video} players. The \texttt{Lenono Video} takes a very significant difference compared to other 4 apps, indicating most of its users are lower-end. One possible reason is the \texttt{Lenovo Video} is a preloaded app that is used mainly on smartphones manufactured by \texttt{Lenovo}, and most of these smartphones are categorized into middle-end and low-end groups. 

Finally, in the browser group, the similar findings can be observed. The most preferred browser of higher-end users is \texttt{Chrome}, followed by the \texttt{360Safe} browser, \texttt{Jinshan} browser, and the \texttt{UC Web} browser. The \texttt{Baidu} browser are the most likely to be adopted by the low-end users.

We can confirm that the device models can affect the choice of competing apps. In addition, we are interested in the possible underlying reasons. Referencing the app similarity model proposed by Chen \textit{et al.}~\cite{Chen:WSDM15}, we explore the profile of these competing apps, including the textual descriptions, vendors, and features. Some immediate findings are derived. 

First, we can infer that the lower-end users prefer the ``local" apps more than the ``international apps". For example, the \texttt{Baidu} browser and the \texttt{360Safe} browser are both developed by local app vendors in China. Second, some special features can be more attractive to different users. For example, from the profile the \texttt{Baidu} and \textit{UC Web} claim that they are more ``traffic-friendly" by compressing the Web page on a front-end proxy before the Web page is delivered to the users. Hence, it is not surprising that these two browsers are more appreciated by the lower-end users, given that these users have limited budgets. Third, the content providers can be an possible impact factor. For example, \texttt{Sohu} is famous for the fast channel of entertainment news in China, and thus the lower-end users, who are possibly in-school students and low-position employees, are more likely to take the \texttt{Sohu News} as the favored app. In contrast, the \texttt{Phoenix News} is provided by the \texttt{Phoenix New Media}\footnote{http://ir.ifeng.com/}, which is famous for the in-depth, objective reviews of economic, finance, and politics. As a result, the users holding higher-end device models the \texttt{Phoenix News} as a preference, since they may care more about the related topics.

\subsection{Summary of Findings}
So far we have answered the 5 research questions proposed previously. Although users holding different device models share similarities in some categories of apps such as \textit{Communication}, significant diversities are observed. The correlation analysis and hypotheses testing confirm that the choice of device models exactly impacts the user behaviors in terms of adoption of app stores, app selection, app abandonment, online time, data plan, app revenue, and selection of competing apps. In summary, we can conclude that the choice of device models can significantly impact the app usage, which in turn reflects the classification of the user requirements, preferences, and possible background. Besides the correlation analysis, we also derive some possible reasons why such diversity exists. We will further explore how our derived knowledge can benefit the research community of mobile computing.

\section{Implications} \label{sec:implication}
In this section, we present some implications and suggestions that might be helpful to relevant stakeholders in the app-centric ecosystem. 

\subsection{App Store Operators}

App stores play the central role in the ecosystem. Intuitively, the recommender systems deployed on the app stores should accurately suggest the proper apps of users. From our findings in \textbf{RQ1}-\textbf{RQ5}, the choice of device models can significantly impact the user preferences of app selection, especially in some competing apps. It is reported that most app stores mainly rely on the similarity-based recommendations such as the apps frequently downloaded by users, the apps developed by the vendors, or apps with similar purposes~\cite{Nicolas:Sigmetrics14}. Some advanced recommendation techniques can further improve the recommendation quality, such as those based on the similarity between apps from various aspects such as app profile, category, permission, images, and updates~\cite{Chen:WSDM15}. However, synthesizing the user requirements, preferences, and even economic background inferred from our study can be further helpful. For example, the app stores can recommend a lower-end user with the browsers such as \texttt{UCWeb} and \texttt{Baidu} instead of \texttt{Chrome}, if the user cares about the data plan. To the best of our knowledge, very few app stores take into account the impacts of device model as an influential factor, including Wandoujia. In practice, we plan to integrate the device model as a dimension to improve the recommendation quality.

From \textbf{RQ1}, we can find that a large number of users (at least in China) do not frequently download and update their apps from app stores. Although the Android users can use other third-party app stores (e.g., those provided by the device manufacturers) other than Wandoujia, it is still worth reporting that the lower-end users are less likely to access the app stores. In this way, the app stores have to carefully consider how to expand the desires from these users.

\subsection{App Developers}
Developers can also learn some lessons from our study when designing and publishing their apps. From \textbf{RQ1}, we can find that some apps are more frequently uninstalled on some device models. It implies that there can be some problems such as compliance with hardware or the device-specific APIs. As reported on the StackOverFlow~\cite{Web:bugs1}, some camera related bugs have been found on \texttt{Samsung Galaxy Note2}. Our finding can validate such problematic issues. When having the distribution of uninstallations according device models, the developers, OS-vendors, and device manufacturers can  draw attentions and prevent problematic issues.

As we presented in \textbf{RQ2}, the choice of device models makes quite significant impacts on the  online time of apps. For example, the higher-end user can spend more time on \textit{NEWS\_AND\_READING} and \textit{TRAVEL}. Due to the fragmentation of Android devices~\cite{khalid2014prioritizing}, currently more and more in-app ads networks allow the developers to customize the title, banner, and content of ads according to the device model~\cite{Web:facebookads}. Hence, the developers of these apps should consider customizing some device-specific ads networks to these ``heavy" users and gain the potential revenue of ads clicking. 

From \textbf{RQ4}, we observe that lower-end users are likely to pay more for apps than higher-end users. Such a finding can assist developers to locate target audiences from whom the revenue can be gained. To attract the lower-end users who are less likely to buy their apps but probably with less budgets, the developers can consider the ``\textit{try-out-and-buy}" model to increase the user interests. In addition, they can further explore some new features or in-app ads to increase more revenue. 

From \textbf{RQ5}, we can find that the selection of competing apps can vary a lot among users holding different levels of device models. Developers can further explore why their apps are less adopted by users from some specific device models, and fix possible bugs, optimize the design, or provide advanced features. For example, it is said that the \texttt{UC Web} can save traffics for users by compressing images and refactoring the Web page layout on the sever before the page is downloaded by the device. Such optimized features can be leveraged to improve the user experiences of apps.  

\subsection{Network Service Providers}
The network service providers or carriers such as T-Mobile, China Mobile, and China UniCom, play an important role in providing the communication infrastructure and service delivery. From \textbf{RQ2} and \textbf{RQ3}, we can derive the network usage patterns of users choosing different device models. From \textbf{RQ2}, the lower-end are more likely to connect to cellular network than the higher-end users, especially in using some specific apps. However, the findings of \textbf{RQ3} suggest that lower-end users typically spend less data plan in cellular network since they may have relatively limited budgets. Such an observation indicates that the lower-end users are less possible to use the ``traffic-heavy" apps such as online music and video players under cellular network. To increase the data plan usage of these lower-end users, the network service providers do need to concern some special business models by case. For example, the network service providers can negotiate with the vendors and provide special packages of data plan, such as the ``\textit{buy-out}" ones that are customized to specific apps, e.g., music or video players. Such packages provide the users the unlimited cellular traffic that can be used but only to access the specific apps. As the customized packages are independent from the regular data plan, they are possibly appreciated by users for specific purposes. The network service providers can exploit such a business model to the device manufacturers and app store operators. For example, the device manufacturers commonly preload some apps in their devices. The network service providers can bind these apps with the customized data package, and share the commissions from the revenue of device manufacturers. 

In addition, from \textbf{RQ2}, we can also infer that the lower-end users are less likely to be covered by Wi-Fi, and tend to spend more network access time under cellular. By synthesizing another findings of \textbf{RQ2}, i.e., the lower-end users spend more time on \textit{EDUCATION} apps under cellular, we can infer that these users are possibly in-school students. As the network service providers can easily obtain the device model information and location distribution of the connected devices, e.g., from the tier-1 cellular carriers, they can estimate and allocate the radio resources around the places where lower-end users are more likely to stay.

\section{Threats to Validity}
Considerable care and attention should be addressed to ensure the rigor of our study. As with any chosen research methodology, it is hardly without limitations. In this section, we will present the potential threats for validating our results.

One potential limitation of our work is that the data set is collected from a single app marketplace in China. The users under study are mainly Chinese, and the region differences should be considered. In addition, the investigated apps are only Android apps. Hence, some results may not be fully generalized to other app stores, platforms, or countries. Such an limitation can hardly be addressed due to the difficulty of acquiring the similar behavioral profile from large-scale users. However, the measurement approach itself can be generalized to other similar data sets from other app stores. Additionally, China has become the biggest market of mobile devices and apps all over the world, so our findings derived from over 0.7 million Android users can provide some comprehensive and representative knowledge to the research communities of modern Internet. 

Choosing the price as indicator can introduce threat since the price of devices can change quite frequently. To address such a threat, we make a coarse-grained categorization of device models. We manually checked the first-to-market price and the latest price as of September 2014 (the end point of our data set) of each model. The price of most device models can still fall into our category. In addition, it would be interesting that if we categorize the device models based on other levels of price. Although there can be some bias caused by the varying price of device models, we believe that our measurement approach and findings can be generalized to any other data sets like ours.
\section{Conclusion And Future Work}

In this paper, we have presented the correlation analysis of choice of device models against the user behaviors of using Android apps. Our study was conducted over the largest to date set collected from over 0.7 million users of Wandoujia. We reported how the choice of device models can impact the adoption of app stores, app selection and abandonment, online time, data plan usage, revenues gained from apps, and comparisons of competing apps. The results revealed the significance of device models against app usage. We summarized our findings and presented implications for relevant stakeholders in the app-centric ecosystem.

Currently, we plan to take into account the device models as an important impact factor in the recommendation systems of Wandoujia~\cite{Liu2009Discovering}. The analytical techniques shall be developed as an offline learning kernel and improve more personalized recommendation of apps.  We are now developing features to collect much finer-grained information in the Wandoujia management apps, such as the traffic/access time per session in apps and the click-through logs. We believe that such detailed information can explore more diversity among users and thus improve our study.



%
%
%
%
%
\balance{}

\balance{}

\bibliographystyle{SIGCHI-Reference-Format2}
\bibliography{price}


\begin{thebibliography}{00}


\ifx \showCODEN    \undefined \def \showCODEN     #1{\unskip}     \fi
\ifx \showDOI      \undefined \def \showDOI       #1{{\tt DOI:}\penalty0{#1}\ }
  \fi
\ifx \showISBNx    \undefined \def \showISBNx     #1{\unskip}     \fi
\ifx \showISBNxiii \undefined \def \showISBNxiii  #1{\unskip}     \fi
\ifx \showISSN     \undefined \def \showISSN      #1{\unskip}     \fi
\ifx \showLCCN     \undefined \def \showLCCN      #1{\unskip}     \fi
\ifx \shownote     \undefined \def \shownote      #1{#1}          \fi
\ifx \showarticletitle \undefined \def \showarticletitle #1{#1}   \fi
\ifx \showURL      \undefined \def \showURL       #1{#1}          \fi

\bibitem{Alharbi:MobileHCI15}
{Khalid Alharbi} {and} {Tom Yeh}. 2015.
\newblock \showarticletitle{Collect, Decompile, Extract, Stats, and Diff:
  Mining Design Pattern Changes in Android Apps}. In {\em Proc. MobileHCI'15}.
  515--524.
\newblock


\bibitem{Matthias:MobileHCI11}
{Matthias B{\"{o}}hmer}, {Brent Hecht}, {Johannes Sch{\"{o}}ning}, {Antonio
  Kr{\"{u}}ger}, {and} {Gernot Bauer}. 2011.
\newblock \showarticletitle{Falling asleep with Angry Birds, {Facebook} and
  {Kindle}: a large scale study on mobile application usage}. In {\em Proc.
  MobileHCI'11}. 47--56.
\newblock


\bibitem{Matthias:CHI13}
{Matthias B{\"{o}}hmer} {and} {Antonio Kr{\"{u}}ger}. 2013.
\newblock \showarticletitle{A study on icon arrangement by smartphone users}.
  In {\em Proc. CHI'13}. 2137--2146.
\newblock


\bibitem{Chen:WSDM15}
{Ning Chen}, {Steven C.~H. Hoi}, {Shaohua Li}, {and} {Xiaokui Xiao}. 2015.
\newblock \showarticletitle{SimApp: {A} Framework for Detecting Similar Mobile
  Applications by Online Kernel Learning}. In {\em Proc. WSDM'15}. 305--314.
\newblock


\bibitem{Church:MobileHCI15}
{Karen Church}, {Denzil Ferreira}, {Nikola Banovic}, {and} {Kent Lyons}. 2015.
\newblock \showarticletitle{Understanding the Challenges of Mobile Phone Usage
  Data}. In {\em Proc. MobileHCI'15}. 504--514.
\newblock


\bibitem{Web:facebookads}
{Facebook}. 2014.
\newblock Facebook: How do I run ads only on specific types of phones?
\newblock   (2014).
\newblock
\newblock
\shownote{\url{https://www.facebook.com/business/help/607254282620194}.}


\bibitem{Falaki:MobiSys10}
{Hossein Falaki}, {Ratul Mahajan}, {Srikanth Kandula}, {Dimitrios
  Lymberopoulos}, {Ramesh Govindan}, {and} {Deborah Estrin}. 2010.
\newblock \showarticletitle{Diversity in smartphone usage}. In {\em Proc.
  MobiSys'10}. 179--194.
\newblock


\bibitem{book:ISMDA}
{Rudolf~J. Freund} {and} {Donna Mohr}. 2010.
\newblock {\em Statistical Methods, Third Edition}.
\newblock {Academic Press}.
\newblock
\showISBNx{978-0123749703}


\bibitem{TSC15Huang}
{Gang Huang}, {Yun Ma}, {Xuanzhe Liu}, {Yuchong Luo}, {Xuan Lu}, {and}
  {M.~Brian Blake}. 2015.
\newblock \showarticletitle{Model-Based Automated Navigation and Composition of
  Complex Service Mashups}.
\newblock {\em {IEEE} Trans. Services Computing\/} {8}, 3 (2015), 494--506.
\newblock


\bibitem{Kelly:THMS13}
{Daniel Kelly}, {Barry Smyth}, {and} {Brian Caulfield}. 2013.
\newblock \showarticletitle{Uncovering Measurements of Social and Demographic
  Behavior From Smartphone Location Data}.
\newblock {\em {IEEE} T. Human-Machine Systems\/} {43}, 2 (2013), 188--198.
\newblock


\bibitem{khalid2014prioritizing}
{Hammad Khalid}, {Meiyappan Nagappan}, {Emad Shihab}, {and} {Ahmed~E Hassan}.
  2014.
\newblock \showarticletitle{Prioritizing the devices to test your app on: A
  case study of Android game apps}. In {\em Proc. FSE'14}. 610--620.
\newblock


\bibitem{WWW16Li}
{Huoran Li}, {Wei Ai}, {Xuanzhe Liu}, {Jian Tang}, {Gang Huang}, {Feng Feng},
  {and} {Qiaozhu Mei}. 2016.
\newblock \showarticletitle{Voting with Their Feet: Inferring User Preferences
  from App Management Activities}. In {\em Proceedings of the 25th
  International Conference on World Wide Web, {WWW} 2016, Montreal, Canada,
  April 11 - 15, 2016}. 1351--1362.
\newblock


\bibitem{Li:IMC15}
{Huoran Li}, {Xuan Lu}, {Xuanzhe Liu}, {Tao Xie}, {Kaigui Bian}, {Felix~Xiaozhu
  Lin}, {Qiaozhu Mei}, {and} {Feng Feng}. 2015.
\newblock \showarticletitle{Characterizing Smartphone Usage Patterns from
  Millions of Android Users}. In {\em Proc. IMC'15}. 459--472.
\newblock


\bibitem{Lim:TSE15}
{Soo~Ling Lim}, {Peter~J. Bentley}, {Natalie Kanakam}, {Fuyuki Ishikawa}, {and}
  {Shinichi Honiden}. 2015.
\newblock \showarticletitle{Investigating Country Differences in Mobile App
  User Behavior and Challenges for Software Engineering}.
\newblock {\em {IEEE} Trans. Software Eng.\/} {41}, 1 (2015), 40--64.
\newblock


\bibitem{TOIS17Liu}
{Xuanzhe Liu}, {Wei Ai}, {Huoran Li}, {Jian Tang}, {Gang Huang}, {Feng Feng},
  {and} {Qiaozhu Mei}. 2017.
\newblock \showarticletitle{Deriving User Preferences of Mobile Apps from Their
  Management Activities}.
\newblock {\em ACM Transactions on Information Systems (TOIS)\/} {35}, 4
  (2017), 39.
\newblock


\bibitem{Liu2009Discovering}
{Xuanzhe Liu}, {Gang Huang}, {and} {Hong Mei}. 2009.
\newblock \showarticletitle{Discovering Homogeneous Web Service Community in
  the User-Centric Web Environment}.
\newblock {\em IEEE Transactions on Services Computing\/} {2}, 2 (2009),
  167--181.
\newblock


\bibitem{Chinaf14Liu}
{Xuanzhe Liu}, {Gang Huang}, {Qi Zhao}, {Hong Mei}, {and} {M.~Brian Blake}.
  2014.
\newblock \showarticletitle{iMashup: A Mashup-Based Framework for Service
  Composition}.
\newblock {\em {SCIENCE} {CHINA} Information Sciences\/} {57}, 1 (2014), 1--20.
\newblock


\bibitem{TSE17Liu}
{Xuanzhe Liu}, {Huoran Li}, {Xuan Lu}, {Tao Xie}, {Qiaozhu Mei}, {Hong Mei},
  {and} {Feng Feng}. 2017.
\newblock \showarticletitle{Understanding Diverse Smartphone Usage Patterns
  from Large-Scale Appstore-Service Profiles}.
\newblock {\em IEEE Transactions on Software Engineering\/} (2017), Accepted to
  appear.
\newblock


\bibitem{TSC15Liu}
{Xuanzhe Liu}, {Yun Ma}, {Gang Huang}, {Junfeng Zhao}, {Hong Mei}, {and}
  {Yunxin Liu}. 2015.
\newblock \showarticletitle{Data-Driven Composition for Service-Oriented
  Situational Web Applications}.
\newblock {\em {IEEE} Trans. Services Computing\/} {8}, 1 (2015), 2--16.
\newblock


\bibitem{ICSE16Lu}
{Xuan Lu}, {Xuanzhe Liu}, {Huoran Li}, {Tao Xie}, {Qiaozhu Mei}, {Dan Hao},
  {Gang Huang}, {and} {Feng Feng}. 2016.
\newblock \showarticletitle{{PRADA:} Prioritizing Android Devices for Apps by
  Mining Large-Scale Usage Data}. In {\em Proceedings of the 38th International
  Conference on Software Engineering, {ICSE} 2016, Austin, TX, USA, May 14-22,
  2016}. 3--13.
\newblock


\bibitem{Ma:WWW2012}
{Haiping Ma}, {Huanhuan Cao}, {Qiang Yang}, {Enhong Chen}, {and} {Jilei Tian}.
  2012.
\newblock \showarticletitle{A habit mining approach for discovering similar
  mobile users}. In {\em Proc. WWW'12}. 231--240.
\newblock


\bibitem{SOSE13Ma}
{Yun Ma}, {Xuanzhe Liu}, {Yihan Wu}, {and} {Paul Grace}. 2013.
\newblock \showarticletitle{Model-Based Management of Service Composition}. In
  {\em Seventh {IEEE} International Symposium on Service-Oriented System
  Engineering, {SOSE} 2013, San Francisco, CA, USA, March 25-28, 2013}.
  103--112.
\newblock


\bibitem{Patro:CoNEXT13}
{Ashish Patro}, {Shravan~K. Rayanchu}, {Michael Griepentrog}, {Yadi Ma}, {and}
  {Suman Banerjee}. 2013.
\newblock \showarticletitle{Capturing mobile experience in the wild: a tale of
  two apps}. In {\em Proc. CoNEXT'13}. 199--210.
\newblock


\bibitem{Rahmati:MobileHCI12}
{Ahmad Rahmati}, {Chad Tossell}, {Clayton Shepard}, {Philip~T. Kortum}, {and}
  {Lin Zhong}. 2012.
\newblock \showarticletitle{Exploring iPhone usage: the influence of
  socioeconomic differences on smartphone adoption, usage and usability}. In
  {\em Proc. MobileHCI'15}. 11--20.
\newblock


\bibitem{Rahmati:TMC13}
{Ahmad Rahmati} {and} {Lin Zhong}. 2013.
\newblock \showarticletitle{Studying Smartphone Usage: Lessons from a
  Four-Month Field Study}.
\newblock {\em {IEEE} Trans. Mob. Comput.\/} {12}, 7 (2013), 1417--1427.
\newblock


\bibitem{Raptis:MobileHCI13}
{Dimitrios Raptis}, {Nikolaos~K. Tselios}, {Jesper Kjeldskov}, {and} {Mikael~B.
  Skov}. 2013.
\newblock \showarticletitle{Does size matter?: investigating the impact of
  mobile phone screen size on users' perceived usability, effectiveness and
  efficiency}. In {\em Proc. MobileHCI'13}. 127--136.
\newblock


\bibitem{Zhong:SIGMobileComm13}
{Ardalan~Amiri Sani}, {Zhiyong Tan}, {Peter Washington}, {Mira Chen}, {Sharad
  Agarwal}, {Lin Zhong}, {and} {Ming Zhang}. 2013.
\newblock \showarticletitle{The wireless data drain of users, apps, {\&}
  platforms}.
\newblock {\em Mobile Computing and Communications Review\/} {17}, 4 (2013),
  15--28.
\newblock


\bibitem{Song:WWW2013}
{Yang Song}, {Hao Ma}, {Hongning Wang}, {and} {Kuansan Wang}. 2013.
\newblock \showarticletitle{Exploring and exploiting user search behavior on
  mobile and tablet devices to improve search relevance}. In {\em Proc.
  WWW'13}. 1201--1212.
\newblock


\bibitem{Web:bugs1}
{StackOverflow}. 2014.
\newblock Android Camera Fails.
\newblock   (2014).
\newblock
\newblock
\shownote{\url{http://stackoverflow.com/search?q=android+camera+samsung+fail}.}


\bibitem{Zhong:CHI2012}
{Chad Tossell}, {Philip~T. Kortum}, {Ahmad Rahmati}, {Clayton Shepard}, {and}
  {Lin Zhong}. 2012.
\newblock \showarticletitle{Characterizing web use on smartphones}. In {\em
  Proc. CHI'12}. 2769--2778.
\newblock


\bibitem{Nicolas:Sigmetrics14}
{Nicolas Viennot}, {Edward Garcia}, {and} {Jason Nieh}. 2014.
\newblock \showarticletitle{A measurement study of {G}oogle play}. In {\em
  Proc. SIGMETRICS'14}. 221--233.
\newblock


\bibitem{Xu:IMC11}
{Qiang Xu}, {Jeffrey Erman}, {Alexandre Gerber}, {Zhuoqing~Morley Mao},
  {Jeffrey Pang}, {and} {Shobha Venkataraman}. 2011.
\newblock \showarticletitle{Identifying diverse usage behaviors of smartphone
  apps}. In {\em Proc. IMC'11}. 329--344.
\newblock


\bibitem{Yan:Mobisys11}
{Bo Yan} {and} {Guanling Chen}. 2011.
\newblock \showarticletitle{{AppJoy}: personalized mobile application
  discovery}. In {\em Proc. MobiSys'11}. 113--126.
\newblock


\end{thebibliography}

\end{document}